\documentclass[useAMS,usenatbib]{mn2e}

\usepackage{graphicx}
\usepackage{aas_macros}
\usepackage[breaklinks,colorlinks=true,citecolor=black,linkcolor=black,urlcolor=blue,bookmarks=true]{hyperref}

\title[Galaxy groups in the Local universe]{
Galaxy groups and clouds in the Local ($z\sim 0.01$) universe
}

\author[Makarov, Karachentsev]{
Dmitry Makarov$^{1,2,3}$\thanks{E-mail: dim@sao.ru},
Igor Karachentsev$^{1,2}$\\
$^{1}$Special Astrophysical Observatory, Nizhniy Arkhyz, Karachai-Cherkessia 369167, Russia\\
$^{2}$Universit\'e de Lyon, Universit\'e Lyon 1, CNRS/IN2P3, Institut de Physique Nucl\'eaire de Lyon, Villeurbanne, France\\
$^{3}$Isaac Newton Institute of Chile, SAO Branch, Russia
}

\begin{document}

\date{Accepted 2010 November 23. Received 2010 November 23; in original form 2010 August 01}
\pagerange{\pageref{firstpage}--\pageref{lastpage}} \pubyear{XXX}
\maketitle

\label{firstpage}

\begin{abstract}
We present an all-sky catalogue of 395 nearby galaxy groups revealed
in the Local Supercluster and its surroundings. The groups and
their associations are identified among 10914 galaxies at 
$|b| > 15^{\circ}$ with radial velocities 
$V_{LG} < 3500$ km s$^{-1}$. 
Our group finding algorithm requires the group members
to be located  inside their zero-velocity surface. Hereby, we
assume that individual galaxy masses are proportional to their
total $K$-band luminosities, $M/L_K = 6\, M_{\odot}/L_{\odot}$.

The sample of our groups, where each group has $n\geq 4$ members,
is characterized by the following medians: mean projected radius $\langle R\rangle=268$ kpc, 
radial velocity dispersion $\sigma_V=74$ km s$^{-1}$, 
$K$-band luminosity $L_K=1.2\,10^{11}$ $L_{\odot}$, 
virial and projected masses $M_{vir}=2.4\,10^{12}$ and $M_{p}=3.3\,10^{12}$ $M_{\odot}$, respectively. 
Accounting for measurement error reduces the median masses by 30 per cent.
For 97 per cent of identified groups the crossing time does not
exceed the cosmic time, 13.7 Gyr, having the median at 3.8 Gyr.

We examine different properties of the groups, in particular, 
of the known nearby groups and clusters in Virgo and Fornax. 
About a
quarter of our groups can be classified as fossil groups where the
dominant galaxy is at least ten times brighter than the other
group members.

In total, our algorithm identifies 54 per cent of galaxies to be members of groups. 
Together with triple systems and pairs they gather 82 per cent of the $K$-band light in Local universe. 
We have obtained the local value of matter density 
to be $\Omega_m = 0.08\pm0.02$ within a distance of $\sim40$ Mpc 
assuming $H_0 = 73$ km s$^{-1}$ Mpc$^{-1}$. 
It is significantly smaller than the cosmic value, 0.28, in the standard $\lambda$CDM model. 
The discrepancy between
the global and local quantities of $\Omega_m$ may be caused by the
existence of Dark Matter component unrelated to
the virial masses of galaxy systems.
\end{abstract}


\section{Introduction}

As the observational data show, the bulk of galaxies inhabit the
groups with a number of members from two to a hundred or more. Our
Galaxy and its companions are no exception, forming a group with
the population $n\sim 25$. The main features of the Local Group
and other closest (and therefore the most studied) groups were
examined by \citet{K2005}. Due to their abundance, the
groups of galaxies make a main contribution to the average
density of matter in the universe. However, according to
\citet{K2005}, this contribution in the Local Volume with
the radius of 10 Mpc around us amounts to just 
$\Omega_{m,{\rm loc}}\sim0.1$  in the units of critical density, 
what is significantly
lower than the global cosmic value $\Omega_{m}\sim 0.28$ 
\citep{FukugitaPeebles2004,Spergel+2007} with the Hubble constant
$H_0=73$ km s$^{-1}$ Mpc$^{-1}$. Such a difference may be due to
the smallness of the Local Volume, where the statistics
of groups is insufficient or does not cover all the variety of groups
according to their morphological population and structure.
Therefore it is essential to determine the mean local density of matter
in larger volume where statistical fluctuations do not introduce 
significant uncertainty.
Despite of 15 per cent variance of density on scale of 80 Mpc \citep{PS2010},
this volume is big enough to be considered as fair approximation 
to mean properties of the Universe.

New mass surveys of galaxy redshifts: 2dF \citep{2dF},
HIPASS \citep{HIPASS}, 6dF \citep{6dF}, ALFALFA
\citep{ALFALFA}, SDSS \citep{SDSS7} present
extensive opportunities for finding the groups of galaxies with a
particular algorithm. However, the surveys of the sky in the
selected regions up to high redshifts $z > 0.1$ appear to be
insufficient for the analysis of small-scale clustering due to the
loss of a great number of low-luminosity dwarf galaxies in the
distant volumes. For example, in the Sloan Digital Sky Survey
(SDSS), covering  a quarter of the sky, the average distance
between the galaxies with measured radial velocities is 9 Mpc,
what is an order of magnitude larger than the diameter of a
typical galaxy group. For a comparison, note that in the
well-studied Local Volume the density of galaxies with measured
velocities is two orders of magnitude higher than in the SDSS.
Therefore, a sensible strategy in the study of galaxy groups would
be to create a representative catalogue of nearby systems over the
entire sky within the radius $z\sim 0.01$.

Successful attempts to create a catalogue of nearby groups were made
by \citet{Vennik1984,Vennik1987,Tully1987,NGC,Magtesyan1988},
who used the method of a `hierarchical tree' proposed by 
\citet{Materne1978,Materne1979}. 
Tully's catalogue and atlas of galaxy groups has 179
pairs and groups, selected among 2367 galaxies with the radial
velocities of less than 3000 km s$^{-1}$. About 2/3 of the 2367
galaxies in the above catalogue appeared to be the members of
multiple systems. Based on the virial mass estimates for these
groups, Tully determined the lower limit of the mean density in
the studied volume as $\Omega_{m,{\rm loc}}\simeq 0.08$. Similar
estimates, $\Omega_{m,{\rm loc}}\sim 0.08$ and 0.05, were obtained by
\citet{Vennik1987} and \citet{Magtesyan1988}, respectively. However, other
authors, \citet{HuchraGeller1982,MdCL1989}, who used the
percolation method (the so-called `friend of friend' method) to
isolate the groups, obtained  3--5 times higher estimates of
$\Omega_m$.

Over the past 20 years the number of galaxies in the volume of the
Local Supercluster and its environs with radial velocities
relative to the centroid of the Local Group $V_{LG}<3500$ 
km s$^{-1}$ has grown by more than 4 times. The updates of the
observational database on the radial velocities, and appearance of
a homogeneous across the sky 2MASS near-infrared photometric survey
\citep{2MASSAtlas,2MASSX} enables us to study the structure and
kinematics of nearby galaxy groups with significantly greater
detail.

This work is a continuation of a series of papers addressing the
properties of binary \citep{Pairs} and triple
\citep{Triplets} systems of galaxies, detected with
one and the same algorithm, applied to the same set of the
observational data. These catalogues contain, respectively, 509
binary systems and 168 triplets of galaxies. In addition to these,
we have compiled a catalogue of 513 isolated galaxies, which are the
most isolated objects in the studied volume $V_{LG}<3500$ km s$^{-1}$ 
\citep{KMKM2009}. In this paper we present a
catalogue of 395 multiple systems with the populations of four or
more members, and discuss the basic properties of these groups.

\section{Observational data}
\label{sec:data}

We use the HyperLEDA\footnote{\url{http://leda.univ-lyon1.fr}}
\citep{HyperLEDA} and 
the NED\footnote{\url{http://nedwww.ipac.caltech.edu}} databases
as main sources of data on radial velocities, apparent
magnitudes, morphological types and other parameters of galaxies.
It must be emphasized that their use requires a critical approach. 
Both these databases contain a significant amount of `spam'.
Quite common case is a misidentification of objects due to 
misprints or imprecise coordinates. 
Without diminishing the importance of mass surveys like 6dF and others,
it is necessary to note they produce significant number of 
erroneous radial velocities.
Apparent magnitudes and radial velocities from the SDSS
survey often correspond to individual knots and associations in
bright galaxies. 
It is only tip of the iceberg of different sources of pollution of the databases.
We have taken into account and corrected, where
possible, these cases, especially significant for selection of
tight galaxy systems. 
As a matter of fact it is most hard and time-consuming part of our work.
Because the databases are constantly updated and new invalid data emerge,
therefore the error correction is iterative task. 
Additionally, we made a number of optical identifications of
$HI$ sources from the HIPASS survey, specifying their coordinates
and determining the apparent magnitudes and morphological types of
galaxies \citep{KMKM2008}. Many dwarf galaxies,
especially of low surface brightness, were examined by us on the
DSS digital images to determine their main characteristics. 
A typical error of our visual estimation of a galaxy's apparent
magnitude is $\sim0.5^m$, and the mean error its type determination is about $\pm2$ 
in the digital scale used by \citet{RC2} in the RC2 catalogue. 

As it is known, the best indicator 
of stellar mass of a galaxy is it's near-infrared luminosity.
The stellar mass dominates the baryon mass in most, but not all, galaxies.
The near-infrared flux is weakly affected by a dust 
and young blue star complexes in the galaxy. 
For this reason, we have taken photometry in $K$-band at $\lambda=2.16\mu m$ as
our photometric basis. Most of these data come from the 
all-sky 2MASS survey \citep{2MASSAtlas,2MASSX}.
In case of lack the $K$-band photometry 
we transferred the optical ($B,V,R,I$) and near-infrared ($J,H$) 
magnitudes of galaxies into the $K$-magnitudes
using synthetic colours of galaxies from 
\citet{Buzzoni2005} and \citet{Fukugita+1995}. 
The greatest amount of photometric data for galaxies
falls on the $B$-band. Based on the relations between the $B-K$
colour index and the morphological type, discussed by 
\citet{2MASSAtlas}, and \citet{KK2005}, we used the
following transformations for the mean colour index:
\begin{eqnarray}
\langle B-K\rangle &\!\!=&\!\! +4.10 \qquad \textrm{for early types, $T\leq2$, (E, S0, Sa),} \nonumber\\
\langle B-K\rangle &\!\!=&\!\! +2.35 \qquad \textrm{where $T\geq9$ (Sm, Im, Irr),} \\
\langle B-K\rangle &\!\!=&\!\! 4.60-0.25 T \qquad \textrm{where $T=3\textrm{--}8$}.\nonumber
\end{eqnarray}
Note that owing to short exposures the 2MASS survey turned out to
be insensitive to the galaxies with low surface brightness and
blue colour. 
For about a thousand dwarf irregular and spheroidal galaxies, 
detected recently by \citet{KK1998,KK2000} in the Local Supercluster volume, 
there are only eye estimate of $B$-magnitudes, 
which converted into $K$-magnitudes using the recipe described above.
Despite the lack of good photometry for them,
gas-rich dIrr galaxies have high-precision radial velocities
from the 21-cm line measurements and 
hence they are important `test particles' to trace
the gravitational well of group of galaxies. 
The need to convert $B$-magnitudes to $K$-band 
for about 35 per cent of galaxies adds a considerable uncertainty 
to the mass estimates of that objects, 
but because we apply it mainly for dwarf galaxies 
which are not dominate in the total luminosity 
this transformation can not change significantly properties of the groups.

We gathered all the available in the HyperLEDA and NED
measurements of radial velocities of the galaxies in the Local
Supercluster and its neighbourhood. Unreliable and inaccurate
measurements, namely, where the velocity measurement error was
greater than 75 km~s$^{-1}$, were excluded. In the data of the
SDSS, 2dF and 6dF surveys we analysed and removed the
measurements with the velocities of $<600$ km~s$^{-1}$, if they
were the cases of a Milky Way star projecting onto a distant
galaxy. When a galaxy had several measurements of its radial
velocity, we chose the median value, and the velocity error was
estimated as a variance of all the good measurements.

It is necessary to note that Local Group with all its known members 
was excluded from calculation because the algorithm does not use
information on real distances and uses only radial velocities for
clusterization. It makes impossible to estimate the properties 
of galaxies in the Local Group that introduces a mess with membership
in the Local Volume.

Our original sample, cleaned from doubtful cases, contains in
total 10914 galaxies with radial velocities in the Local Group
rest frame of $V_{LG}<3500$ km s$^{-1}$, located at the galactic
latitudes $|b|>15^{\circ}$. 
For all these galaxies we fixed their apparent magnitudes and morphological types.
To avoid influence of the boundaries on properties of groups
we also used in our calculations the data on the galaxies
located in the boundary regions with $10^{\circ} <|b| <15^{\circ}$ 
and with $3500<V_{LG}<4000$ km s$^{-1}$, as some
individual members of groups with large virial velocities could
appear there. The sampling of such a depth contains the entire
Local Supercluster with its distant outskirts, surrounding voids
and ridges of the  neighbouring clusters.

\section{The group finding algorithm}

Various algorithms were proposed to identify the groups of
galaxies in samples, limited by apparent magnitudes or galaxy
distances. However, they can be reduced to two basic ones: the
percolation method (`friend of friend') and the taxonometric
method (construction of a hierarchical tree).

Using the percolation technique, \citet{HuchraGeller1982} combined
the galaxies in groups on the condition that their projected
mutual linear separations and radial velocity differences were
smaller than some threshold values $R_c$ and $V_c$. At $R_c=0.52$
Mpc and $V_c=600$ km s$^{-1}$ they have grouped in the CfA
redshift survey about 74 per cent of the galaxies, and obtained the
groups with a characteristic radius of $R_h=1.1$ Mpc, radial
velocity dispersion $\sigma_V=208$ km s$^{-1}$, and mean virial
mass $\lg (M_{vir}/M_{\odot})=13.5$. This method was applied by
many authors to different samples of galaxies. The disadvantage of
this method is the arbitrariness of choice of two percolation
parameters $R_c$ and $V_c$, a variation of which strongly affects
the characteristic size and mass of groups, as well as the
fraction of galaxies belonging to the groups. Tracking some mean
contrast of the galaxy number density by the $R_c$ and $V_c$
parameters, the percolation criterion overlooks many real groups
in the regions of low density, and finds large non-virialized
aggregates in the regions of overdensity. Another shortcoming of
the `friend of friend' method manifests itself in a strong
dependence of the group parameters  on the group distance $D$ from
the observer. Various attempts to reduce this dependence by
introducing the variables $R_c(D)$ and $V_c(D)$ were accompanied
by additional arbitrary assumptions. Recently, \citet{Crook+2007}
applied the percolation method to the 2MASS survey of galaxies,
and identified 1710 pairs and 1258 groups of galaxies at the
relative density contrast $\delta\rho/\rho=80$. In this sample,
the members of groups and pairs make up, respectively, 36 per cent and
17 per cent. The groups by \citet{Crook+2007} with the number of members
$n\geq 5$ have a characteristic projection radius of about 1 Mpc,
the dispersion of radial velocities of $\sim200$ km s$^{-1}$ and
the mean virial mass $\lg(M_{vir}/M_{\odot})\sim13.5$. Taken the
depth of the considered 2MASS sample $D_{max}=140$ Mpc, the
contribution of virial masses of these groups in the mean density
of matter is $\Omega_m \simeq 0.13$.

Following another, `taxonometric' method, \citet{Vennik1984,NGC}
combined galaxy pairs by the maximum ratio of their
luminosity to the cube of mutual distance ($L_{ik}/R_{ik}^3$).
Then such a pair was replaced by a `particle' with the total
luminosity, and the process of finding a case with
$\max(L_{ik}/R_{ik}^3)$ repeated. The process was completed by
creation of a single `hierarchical tree' whose branches united
the entire considered sample of galaxies. Clipping the tree
branches on some contrast level  of the volume luminosity yielded
a set of branches-groups, the sizes and virial masses of which
were dependent on the selected density (luminosity) contrast.
Applying the dendrogram method, both authors obtained a
characteristic projection radius of the group of 0.3 Mpc, the mean
radial velocity dispersion $\sigma_V \simeq 100$km s$^{-1}$, and
the virial mass to blue luminosity ratio 
$M_{vir}/L_B \simeq 95$ $M_{\odot}/L_{\odot}$.

The both percolation and dendrogram methods ignore the
individual properties of galaxies, considering them as
indistinguishable particles. But it is obvious that the same
thresholds $R_c$ and $V_c$ may be sufficient for clustering a
pair of dwarf galaxies, but they are apparently not sufficient to
bind a pair of giant galaxies. This inadequacy of the algorithm
leads to a systematic distortion of the virial mass estimates.

Combining the galaxies into the systems of different multiplicity
should be done taking into account the individual properties of
galaxies. Considering two arbitrary galaxies as a virtual bound
pair, we assume \citep{K1994,MK2000} that the spatial velocity 
difference $V_{12}$ for the
galaxies in a physical pair and their spatial separation $R_{12}$
must satisfy the condition of negative total energy
\begin{equation}
\frac{V_{12}^2R_{12}}{2GM_{12}}<1,
\label{eq:PhysCrit}   
\end{equation}
where $M_{12}$ is the total mass of the pair, and $G$ is the
gravitational constant. However, from the observations we only
know the velocity difference projected on the line of sight
$V_{12,r}$, and the separation projected onto the image plane
$R_{\bot}$. Two galaxies with a very small difference in radial
velocities but a large separation in the sky can satisfy the
condition (\ref{eq:PhysCrit}) without being mutually bound. 
Hence the condition of negative total energy of the pair, 
expressed in terms of the observables
\begin{equation}
\frac{V^2_{12,r}R_{\bot}}{2GM_{12}}<1
\label{eq:Criterion1}
\end{equation}
must be supplemented by another restriction on the maximal
distance between the components at their fixed mass $M_{12}$. The
condition when the pair components remain within the sphere of
`zero-velocity' \citep{Sandage1986} takes the form of
\begin{equation}
 \frac{\pi H_0^2R^3_{\bot}}{8GM_{12}}<1,
\label{eq:Criterion2}
\end{equation}
where $H_0$ is the Hubble constant.

Our algorithm for galaxy grouping is in fact a variant of the
percolation method. 
Firstly, we select all the pairs satisfying the conditions 
(\ref{eq:Criterion1}) and (\ref{eq:Criterion2}).
Then all the pairs with common main component are combined in a group. 
If a galaxy turned out to be a companion of several massive galaxies at once,
we join it with the most massive neighbour. 
In a particular case, one group may be a subgroup within a more extended system. 
In this sense, our algorithm combines the properties of both 
the `friend of friend' method, and the hierarchical approach.
On next stage, we replace the galaxies in the group by fake object 
with summarized luminosity of all its members and with mean redshift.
After that we repeat all the steps from the beginning 
while some object obeys to bound criteria.
Although, the algorithm is based on pairwise criterion 
on final step the bound condition is determined by the entire group.

We determined the masses of galaxies from their integral
luminosity in the near-infrared $K_s$-band, supposing that they have
the same mass-to-luminosity ratio
\begin{equation}
  M/L_K={\kappa} (M_{\odot}/L_{\odot}),
\label{eq:ML}
\end{equation}
where $\kappa$ is taken equal to 6. 
In the fact, the value of $\kappa=6$ is only more or less 
arbitrary dimensionless parameter of the algorithm. 
To bound it we `trained' the clusterization algorithm 
(\ref{eq:Criterion1}--\ref{eq:ML}) on detailed three-dimensional
distribution of galaxies in the Local Volume, where 
the membership of galaxies in the groups is known from good quality
photometric distances. 
\citet{K2005} lists the members of several nearby groups like Cen~A and M~81. 
Unfortunately, as it was noted in previous section,
we can not test the algorithm on Local Group, 
thus we used other nearby groups in the Local Volume.
The choice of $\kappa=6$ is the compromise between a loss of the real members
and an impurity of groups by false members.
For the $\kappa\le4$ we lose significant number of real members
while $\kappa\ge8$ leads to appearance in the groups suspicious members.
Moreover, for $\kappa\ge10$ galaxies are combined into extended 
non-virialized aggregates. 
At the given value of $\kappa=6$ the
dwarf companions in the well-known nearby groups are usually
located inside the zero velocity surface around the major galaxies
of these groups.

\section{The catalogue of groups}

The criteria (\ref{eq:Criterion1}--\ref{eq:ML}) of unifying 
the galaxies in groups with the
parameter $\kappa=6$ was used for 10914 galaxies with radial
velocities $V_{LG}< 3500$ km s$^{-1}$, located outside the Milky
Way zone, $|b| > 15^{\circ}$.  This led to an
identification of 395 groups with the population of $n\geq4$
members. In total, these groups include 4381 galaxies. 
Together with 1018 components of binary systems \citep{Pairs}
and 504 components of triplets \citep{Triplets} in
the same volume, the total number of clustered galaxies is 5903 or
54 per cent of the total number considered.

The catalogue of galaxy groups as the final result of successive
iterations of the use of conditions (\ref{eq:Criterion1}--\ref{eq:ML})
 is presented in a short
and full version. Table~\ref{tab:groups} is a compact version of the catalogue,
containing the basic group characteristics listed in one row. The
full version of the catalogue with the indication of all the
individual members of each group is available in the electronic
form at \url{http://www.sao.ru/hq/dim/groups}.

The columns of Table~\ref{tab:groups} contain the following data:
\par1) principal name of the group's brightest galaxy, taken, as a
rule, from the LEDA;
\par2) equatorial coordinates of the group's main member at the epoch
(J2000.0);
\par3) the number of group members with known radial velocities;
\par4) mean radial velocity of the group in km s$^{-1}$ relative to
the centroid of the Local Group;
\par5) the standard deviation of radial velocities of the group
members (km s$^{-1}$) not corrected for the velocity measurement
errors;
\par6) mean harmonic radius of the group (kpc); at its computation
the distance to the group $\langle D \rangle$ was determined from the mean
radial velocity with the Hubble parameter $H_0=73$ km s$^{-1}$ Mpc$^{-1}$;
\par7) logarithm of the total luminosity of the group in the
photometric $K_s$-band given in the unit of solar luminosity at
$M_{\odot,K} =3.28^m$ \citep{BinneyMerrifield1998};
\par8) logarithm of the projected mass of the group, 
as defined by \citep[][equation 11]{Heisler+1985}
\begin{equation}
M_p= \frac{32}{\pi} \frac{1}{N-3/2} \sum_{i=1}^N \frac{V^2_{i,r} R_{i,\bot}}{G}\nonumber
\label{eq:Mproj}
\end{equation}
where $V_{i,r}$ and $R_{i,\bot}$ are radial velocity and projected distance
of the $i$th galaxy relative to the centre of the system.
It should be noted that this value is statistically biased.
To obtain an unbiased mass estimate the square of velocity, $V_{i,r}^2$, 
has to be corrected for measurement error, $(V_{i,r}^2-\epsilon^2)$. 
In case of large errors, $\epsilon$, 
the unbiased value of the group mass, $M^c_p$, attains a negative value;
\par9) the projected mass-to-total luminosity ratio in the $K$-band in
solar units;
\par10) morphological type of the group's main member according to
the RC2 classification  \citep{RC2};
\par11) difference in apparent $K$-magnitudes of the first and second
members of the group, ranked by $K$-luminosity;
\par12) group's membership in an association (cloud, clan), which is
identified at a higher value of the dimensionless parameter
adopted to be $\kappa = 40$; here the name of the association was
given by the name of the group that is dominant in it; as it is
evident from these data, a significant number of groups are
isolated entities not associated with other  neighbouring groups.

\begin{table*}
\footnotesize
\caption{Main properties of groups.\label{tab:groups}}
\begin{tabular}{lcrrrrrrrrrl}
 Group           &      J2000.0      &   N & $V_{LG}$  & $\sigma_V$ & $R_h$ & $\lg L/L_\odot$ & $\lg M/M_\odot$ & $\lg M/L$ &  T & $\Delta m_{21}$ &    Association \\
\hline\hline
 NGC7814         & 000314.9$+$160844 &   4 & 1283 &   34 & 262 & 10.98 &  12.16 & 1.18 &$  2 $& 5.27 & NGC7814             \\
 ESO293--034     & 000619.9$-$412960 &   5 & 1435 &   59 &  41 & 10.18 &  11.87 & 1.69 &$  6 $& 3.14 & ESO293--034         \\
 NGC0055         & 001453.6$-$391148 &   4 &  111 &   26 & 123 &  9.46 &  11.37 & 1.91 &$  9 $& 0.13 & NGC0055             \\
 NGC0092         & 002131.7$-$483729 &   5 & 3302 &  101 &  40 & 11.08 &  12.49 & 1.41 &$  1 $& 1.16 & NGC0092             \\
 NGC0134         & 003022.0$-$331439 &   4 & 1629 &  121 & 159 & 11.29 &  12.54 & 1.25 &$  4 $& 3.64 & NGC0134             \\
 NGC0253         & 004733.1$-$251718 &   6 &  374 &   87 & 275 & 11.24 &  12.87 & 1.63 &$  5 $& 3.66 & NGC0253             \\
 NGC0289         & 005242.4$-$311221 &   8 & 1618 &   81 & 205 & 11.01 &  12.77 & 1.76 &$  4 $& 0.73 & NGC0134             \\
\hline
\end{tabular}
\end{table*}

\addtocounter{table}{-1}
\begin{table*}
\footnotesize
\caption{Continue}
\begin{tabular}{lcrrrrrrrrrl}
 Group           &      J2000.0      &   N & $V_{LG}$  & $\sigma_V$ & $R_h$ & $\lg L/L_\odot$ & $\lg M/M_\odot$ & $\lg M/L$ &  T & $\Delta m_{21}$ &    Association \\
\hline\hline
 NGC2300         & 073220.5$+$854232 &  11 & 2100 &   85 & 409 & 11.45 &  12.86 & 1.41 &$ -3 $& 1.49 & NGC2300             \\
 NGC2442         & 073623.8$-$693151 &  12 & 1135 &   63 & 150 & 11.24 &  12.24 & 1.00 &$  4 $& 1.00 & NGC2442             \\
 NGC2565         & 081948.3$+$220153 &   5 & 3467 &   51 & 113 & 11.15 &  12.26 & 1.11 &$  4 $& 0.49 & NGC2565             \\
 NGC2577         & 082243.5$+$223311 &   4 & 2024 &  108 & 129 & 10.82 &  12.30 & 1.48 &$ -3 $& 0.46 & NGC2577             \\
 NGC2551         & 082450.3$+$732443 &   5 & 2436 &   51 & 301 & 10.78 &  12.09 & 1.31 &$  0 $& 1.03 & NGC2633             \\
 NGC2594         & 082717.2$+$255244 &   4 & 2215 &   55 & 124 & 10.18 &  11.69 & 1.51 &$  1 $& 3.03 & NGC2592             \\
 NGC2604A        & 083323.1$+$293220 &   4 & 2013 &   25 &   8 &  9.85 &  10.98 & 1.13 &$  6 $& 2.04 & NGC2608             \\
\hline                                                                                  
\end{tabular}
\end{table*}

\addtocounter{table}{-1}
\begin{table*}
\footnotesize
\caption{Continue}
\begin{tabular}{lcrrrrrrrrrl}
 Group           &      J2000.0      &   N & $V_{LG}$  & $\sigma_V$ & $R_h$ & $\lg L/L_\odot$ & $\lg M/M_\odot$ & $\lg M/L$ &  T & $\Delta m_{21}$ &    Association \\
\hline\hline
 NGC3230         & 102344.0$+$123404 &   6 & 2728 &  134 & 111 & 10.94 &  12.49 & 1.55 &$  1 $& 3.61 & NGC3230             \\
 NGC3245         & 102718.4$+$283027 &   6 & 1291 &   39 & 239 & 10.95 &  11.91 & 0.96 &$ -2 $& 0.94 & NGC3245             \\
 ESO375--041     & 102931.0$-$351535 &   4 & 1476 &   39 &  58 & 10.07 &  11.16 & 1.09 &$ -2 $& 0.64 & NGC3223             \\
 NGC3275         & 103051.8$-$364413 &   5 & 2852 &   40 & 674 & 11.29 &  12.44 & 1.15 &$  2 $& 1.64 & NGC3223             \\
 IC2587          & 103059.6$-$343347 &   6 & 1701 &   91 & 225 & 10.58 &  12.38 & 1.80 &$ -3 $& 1.77 & NGC3223             \\
 NGC3281         & 103152.1$-$345113 &   4 & 3116 &  135 & 510 & 11.35 &  13.31 & 1.96 &$  2 $& 1.72 & NGC3223             \\
 NGC3282         & 103221.9$-$221808 &   4 & 3416 &   72 & 147 & 10.84 &  12.17 & 1.33 &$ -2 $& 3.50 & NGC3311             \\
\hline
\end{tabular}
\end{table*}

\addtocounter{table}{-1}
\begin{table*}
\footnotesize
\caption{Continue}
\begin{tabular}{lcrrrrrrrrrl}
 Group           &      J2000.0      &   N & $V_{LG}$  & $\sigma_V$ & $R_h$ & $\lg L/L_\odot$ & $\lg M/M_\odot$ & $\lg M/L$ &  T & $\Delta m_{21}$ &    Association \\
\hline\hline
 NGC4030         & 120023.6$-$010600 &   7 & 1291 &   50 & 347 & 10.91 &  12.52 & 1.61 &$  4 $& 3.57 & NGC4472             \\
 NGC4039         & 120153.6$-$185311 &  23 & 1409 &   74 & 256 & 11.50 &  12.82 & 1.32 &$  9 $& 0.12 & NGC4039             \\
 UGC07017        & 120222.5$+$295142 &   4 & 3100 &   44 & 123 & 10.56 &  11.70 & 1.14 &$  4 $& 1.35 & NGC3966             \\
 NGC4062         & 120403.8$+$315345 &   4 &  736 &   33 & 175 & 10.08 &  11.92 & 1.84 &$  5 $& 2.55 & NGC4472             \\
 NGC4105         & 120640.8$-$294537 &  29 & 1870 &  139 & 382 & 11.61 &  13.23 & 1.62 &$ -4 $& 0.60 & NGC3923             \\
 NGC4111         & 120703.1$+$430355 &  20 &  851 &   93 & 212 & 11.14 &  12.69 & 1.55 &$ -1 $& 0.08 & NGC5194             \\
 UGCA 272        & 120747.6$+$672302 &   4 & 2441 &   33 & 174 & 10.18 &  11.50 & 1.32 &$  2 $& 1.52 & NGC4256             \\
\hline
\end{tabular}
\end{table*}

\addtocounter{table}{-1}
\begin{table*}
\footnotesize
\caption{Continue}
\begin{tabular}{lcrrrrrrrrrl}
 Group           &      J2000.0      &   N & $V_{LG}$  & $\sigma_V$ & $R_h$ & $\lg L/L_\odot$ & $\lg M/M_\odot$ & $\lg M/L$ &  T & $\Delta m_{21}$ &    Association \\
\hline\hline
 NGC4759         & 125304.5$-$091160 &  12 & 3471 &  146 & 454 & 11.81 &  13.55 & 1.74 &$ -2 $& 0.63 & NGC4759             \\
 NGC4795         & 125502.9$+$080356 &   7 & 2548 &  118 &  53 & 10.80 &  12.34 & 1.54 &$  1 $& 0.78 & NGC4472             \\
 NGC4814         & 125521.9$+$582039 &   5 & 2644 &   76 &  78 & 10.91 &  12.63 & 1.72 &$  3 $& 2.03 & NGC4814             \\
 NGC4808         & 125549.0$+$041815 &   5 &  591 &   27 & 115 &  9.78 &  11.30 & 1.52 &$  6 $& 0.72 & NGC4472             \\
 NGC4835         & 125807.8$-$461551 &   5 & 1933 &   59 &  83 & 10.92 &  12.31 & 1.39 &$  4 $& 4.84 & NGC4696             \\
 NGC4856         & 125921.3$-$150232 &   5 & 1189 &   50 & 217 & 10.79 &  12.25 & 1.46 &$ -0 $& 3.58 & NGC4472             \\
 NGC4866         & 125927.1$+$141016 &   5 & 1909 &   58 & 126 & 11.02 &  12.63 & 1.61 &$ -0 $& 3.65 & NGC4472             \\
\hline
\end{tabular}
\end{table*}

\addtocounter{table}{-1}
\begin{table*}
\footnotesize
\caption{Continue}
\begin{tabular}{lcrrrrrrrrrl}
 Group           &      J2000.0      &   N & $V_{LG}$  & $\sigma_V$ & $R_h$ & $\lg L/L_\odot$ & $\lg M/M_\odot$ & $\lg M/L$ &  T & $\Delta m_{21}$ &    Association \\
\hline\hline
 NGC5643         & 143240.8$-$441029 &   4 &  986 &   59 & 167 & 10.92 &  12.48 & 1.56 &$  5 $& 0.70 & NGC5643             \\
 NGC5676         & 143246.8$+$492728 &  17 & 2389 &  105 & 348 & 11.57 &  12.99 & 1.42 &$  5 $& 0.48 & NGC5676             \\
 NGC5707         & 143730.8$+$513343 &   7 & 2348 &   87 & 220 & 10.73 &  12.12 & 1.39 &$  2 $& 2.26 & NGC5676             \\
 UGC09476        & 144132.0$+$443046 &   4 & 3401 &   23 & 100 & 10.77 &  11.76 & 0.99 &$  5 $& 0.56 & UGC09476            \\
 NGC5729         & 144206.9$-$090034 &   4 & 1750 &   49 & 106 & 10.43 &  11.76 & 1.33 &$  3 $& 1.71 & NGC5729             \\
 NGC5726         & 144256.0$-$182642 &   4 & 3333 &   54 & 134 & 10.81 &  12.02 & 1.21 &$ -3 $& 3.29 & NGC5726             \\
 IC1048          & 144258.0$+$045322 &   4 & 1671 &   59 & 132 & 10.27 &  11.98 & 1.71 &$  3 $& 2.42 & NGC5846             \\
\hline
\end{tabular}
\end{table*}

\addtocounter{table}{-1}
\begin{table*}[t]
\footnotesize
\caption{Continue}
\begin{tabular}{lcrrrrrrrrrl}
 Group           &      J2000.0      &   N & $V_{LG}$  & $\sigma_V$ & $R_h$ & $\lg L/L_\odot$ & $\lg M/M_\odot$ & $\lg M/L$ &  T & $\Delta m_{21}$ &    Association \\
\hline\hline
 IC1459          & 225710.6$-$362744 &  16 & 1775 &  221 & 336 & 11.66 &  13.54 & 1.88 &$ -5 $& 1.57 & NGC7582             \\
 IC5267          & 225713.6$-$432346 &   5 & 1644 &   93 & 185 & 11.21 &  12.48 & 1.27 &$ -1 $& 1.14 & NGC7582             \\
 UGC12281        & 225912.8$+$133624 &   4 & 2871 &   39 &  85 &  9.73 &  11.18 & 1.45 &$  7 $& 2.33 & UGC12281            \\
 NGC7454         & 230107.3$+$162258 &   7 & 2248 &   83 &  86 & 10.85 &  12.46 & 1.61 &$ -5 $& 0.06 & NGC7454             \\
 NGC7484         & 230704.9$-$361631 &   4 & 2707 &  103 &  40 & 10.95 &  12.37 & 1.42 &$ -5 $& 3.38 & NGC7484             \\
 NGC7507         & 231207.6$-$283223 &   4 & 1605 &   80 &  57 & 11.18 &  12.00 & 0.82 &$ -5 $& 1.67 & NGC7507             \\
 NGC7582         & 231823.5$-$422214 &  13 & 1578 &   67 & 299 & 11.64 &  12.61 & 0.97 &$  2 $& 0.22 & NGC7582             \\
 IC5328          & 233316.5$-$450057 &   6 & 3084 &   67 & 566 & 11.29 &  12.82 & 1.53 &$ -4 $& 3.52 & IC5328              \\
 NGC7716         & 233631.5$+$001750 &   5 & 2797 &   40 & 291 & 10.81 &  11.79 & 0.98 &$  3 $& 2.49 & NGC7716             \\
 NGC7727         & 233953.9$-$121735 &   5 & 2043 &   51 & 314 & 11.37 &  12.19 & 0.82 &$  1 $& 0.59 & NGC7727             \\
 NGC7743         & 234421.1$+$095603 &   4 & 1856 &   81 &  94 & 11.03 &  12.29 & 1.26 &$ -1 $& 0.22 & NGC7743             \\
\hline\hline
\end{tabular}
\end{table*}

\section{Basic properties of the groups}

\begin{figure*}
\caption{
Sky distribution of the groups in equatorial coordinates. 
The groups are plotted as circles with a diameter proportional to the group $K$-band luminosity. 
The circle colour indicates a morphological type of the main galaxy in the group. 
The upper, middle and bottom panels corresponds, respectively, 
to three different volumes separated by the mean radial velocity of the groups. 
Small dots present the distribution of individual galaxies with radial velocities in the above intervals. 
The zone of strong Galactic extinction is shown in grey.
}
\label{fig:allsky}
\end{figure*}

\begin{figure*}
\caption{
Sky distribution of the galaxy associations identified by the clustering algorithm with $\kappa = 40$. 
The shaded polygons outline real angular sizes of the systems. 
Three panels correspond to the same volumes as in Fig.~\ref{fig:allsky}.
}
\label{fig:assoc}
\end{figure*}

The distribution of groups of galaxies over the sky in equatorial
coordinates is given in three panels of Fig.~\ref{fig:allsky}. 
The upper, middle
and bottom panels correspond to near, intermediate and far
volumes, delimited by the mean radial velocities of the groups:
$V_{LG}< 1200$ km s$^{-1}$, $1200 < V_{LG} < 2400$  km s$^{-1}$
and $2400 < V_{LG} < 3500$  km s$^{-1}$. Each group is indicated
by a circle, the diameter of which is proportional to the total
$K$-luminosity (i.e. the stellar mass) of the group members. The
circle colour indicates the morphological type of the main member
of the group in the colour spectrum: from the early E, S0 types
with an old population (red) to the late Irr, BCD types with a
young population (light blue). For comparison, small black dots
show the distribution of individual galaxies with radial
velocities in the above intervals. The grey ragged ring-shaped
region traces the zone of strong absorption in the Milky Way
according to \citet{Dustmap}.

As expected, a complex of groups in the Virgo region, where the
core of the Local Supercluster of galaxies is located, stands
out in the near volume ($V_{LG} < 1200 $km s$^{-1}$). The richest
group ($n=355$ members) is associated with a giant elliptical
galaxy M~49 = NGC~4472. The  neighbouring groups with the total
number of members amounting to 1558 are located mostly along the
equator of the Local Supercluster. The other most massive groups
in this volume are the groups around M~105 = NGC~3379 (Leo~I) and
NGC~1553 (Dorado). Among the nearest groups our criteria
identifies practically all the known groups around the main
galaxies: M~81, NGC~253, Cen~A, M~83, NGC~628, M~51, M~101,
NGC~891/1023, M~104 (Sombrero) etc.

In the intermediate volume ($1200 < V_{LG} < 2400$) the traces of
group concentration within the Virgo region are also visible, but
in general the effect of the Local Supercluster is barely
noticeable. The most massive group in this volume is the
Fornax cluster (NGC~1399) with a population of $n=111$.
The aggregate of groups in the Fornax and in the Eridanus 
(NGC~1332/1395/1407) gathers 379 galaxies in total.
In addition to these, in
other regions of the sky there are massive groups around NGC~5846
$(n=74)$ and NGC~5746 $(n=39)$.

In the distant part of the studied volume (the bottom panel of
Fig.~\ref{fig:allsky}) the groups show a tendency to be located along some
filaments. The most massive groups: Centaurus (NGC~4696), Antlia
(NGC~3268) and Hydra (NGC~3311) have masses comparable with that
of the Virgo and Fornax clusters. It is easy to notice that the
colour of circles in all the panels is correlated with their sizes,
demonstrating the well-known observational fact that the E and S0
galaxies usually occur among the brightest members of rich groups
and clusters.

Increasing the clustering parameter $\kappa$ in (\ref{eq:ML}), the galaxies
can be combined into more extended aggregates (clouds), which no
longer meet the condition of virial equilibrium. However, the
members of such associations still probably lie within their
common `zero-velocity surface', i.e. they will approach each
other and be subjected to the subsequent virialization. Three
panels in Fig.~\ref{fig:assoc} show the distribution of galaxies in the sky with
radial velocities in the same intervals as in Fig.~\ref{fig:allsky}.
The shaded
polygons there represent the zones of such aggregates with
their real angular sizes in the sky. In the nearby volume (the top
panel) the most extended aggregate is the Virgo cluster, and its
angular diameter is approximately equal to the diameter of the
`zero-velocity surface' of the cluster, which is
$2R_0=46^{\circ}$ according to \citet{KN2010}. In
the middle panel in the volume of $1200 < V_{LG} < 2400$ km s$^{-1}$, 
the most extended aggregate is the Fornax+Eridanus
association of groups. Its angular size is also close to the
diameter $2R_0$, which is equal to  $\simeq 22^{\circ}$ as
estimated by \citet{Nasonova+2011}. These correspondences suggest
that the dimensions of other associations (overdensities) from
Fig.~\ref{fig:assoc} may be also in accordance with sizes of their `infall zones'. 
It will be appropriate to note here that there is a
significant fraction of galaxies that are located outside the
volumes of both groups and clouds. Their distribution does not
appear to be random, but shows some correlation with the
distribution of group centres. The most isolated of these `field galaxies'
\citep{KMKM2009} are of considerable interest with regard to the effect
of  their isolation on the population and structure of these
objects, and the star formation rate in them.

\begin{figure*}
\includegraphics[width=0.8\textwidth]{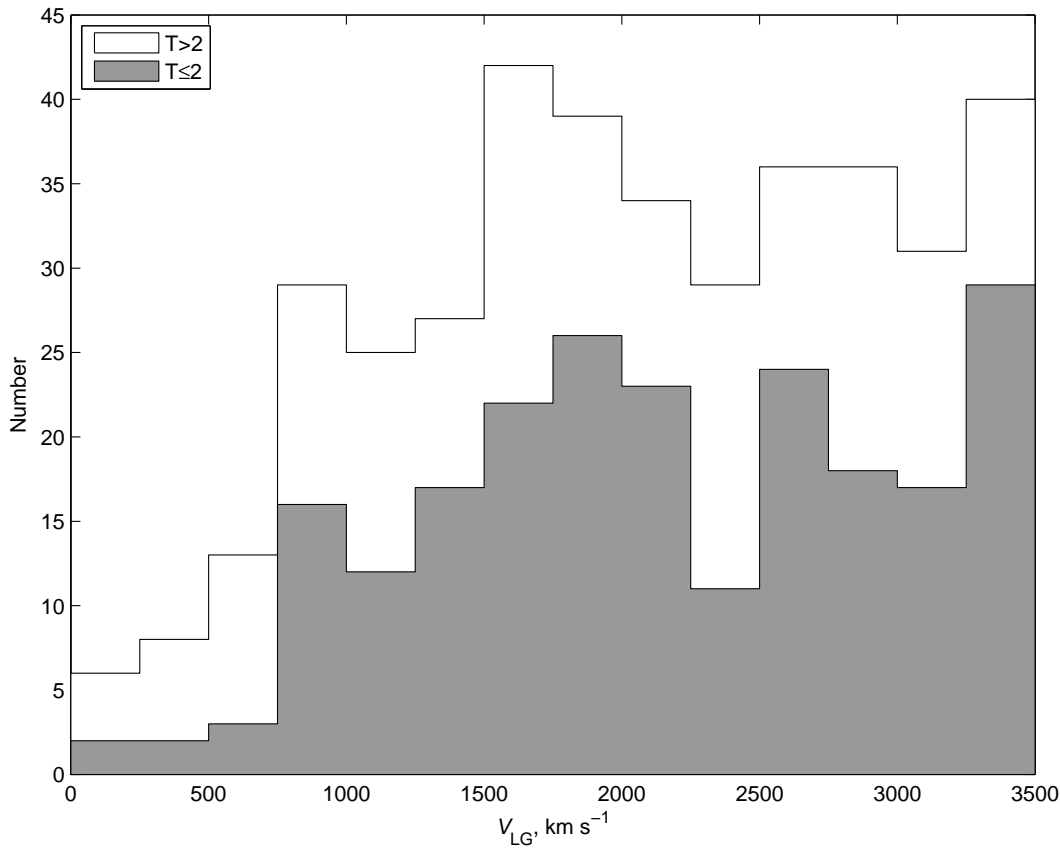}
\caption{
Distribution of 395 identified groups by their mean radial velocities. 
The groups with a bulge-dominated principal galaxy ($T\leq2$) are shown in grey.
}
\label{fig:h_vlg}
\end{figure*}

Among 10914 galaxies in the studied volume, the individual
distance estimates are so far known  for less than 2000 galaxies.
Most of them are spiral galaxies, where the distances were
determined by the \citet{TullyFisher} method with the accuracy of
$\sim20$ per cent from the relation between the luminosity and rotation
amplitude of the galaxies. However, to ensure a uniform approach
we used only the Hubble distances of galaxies $D= V_{LG}/H_0$,
that, of course, distorts the true picture of distribution of
galaxy groups in the Local universe. 
The distribution of 395 identified groups by their mean radial velocities 
is presented in Fig.~\ref{fig:h_vlg}. 
The groups in which the main member is a galaxy with a
well-developed bulge ($T\leq2)$ are marked in the figure in grey.
As one can see, the distribution of $N(V_{LG})$ is markedly
different from homogeneous, showing an excess of the group number
from the Virgo and Fornax complexes with their velocities around
1500 km s$^{-1}$. In the Local Volume ($V_{LG}< 750$ km s$^{-1}$)
there exists a lack of groups with the main galaxies of early type.
It reflects the well known effect of the morphological segregation 
with environment because our Galaxy lies on the edge of 
Local Supercluster far away from 
the dense concentration of the galaxies.

\begin{figure*}
\includegraphics[width=0.8\textwidth]{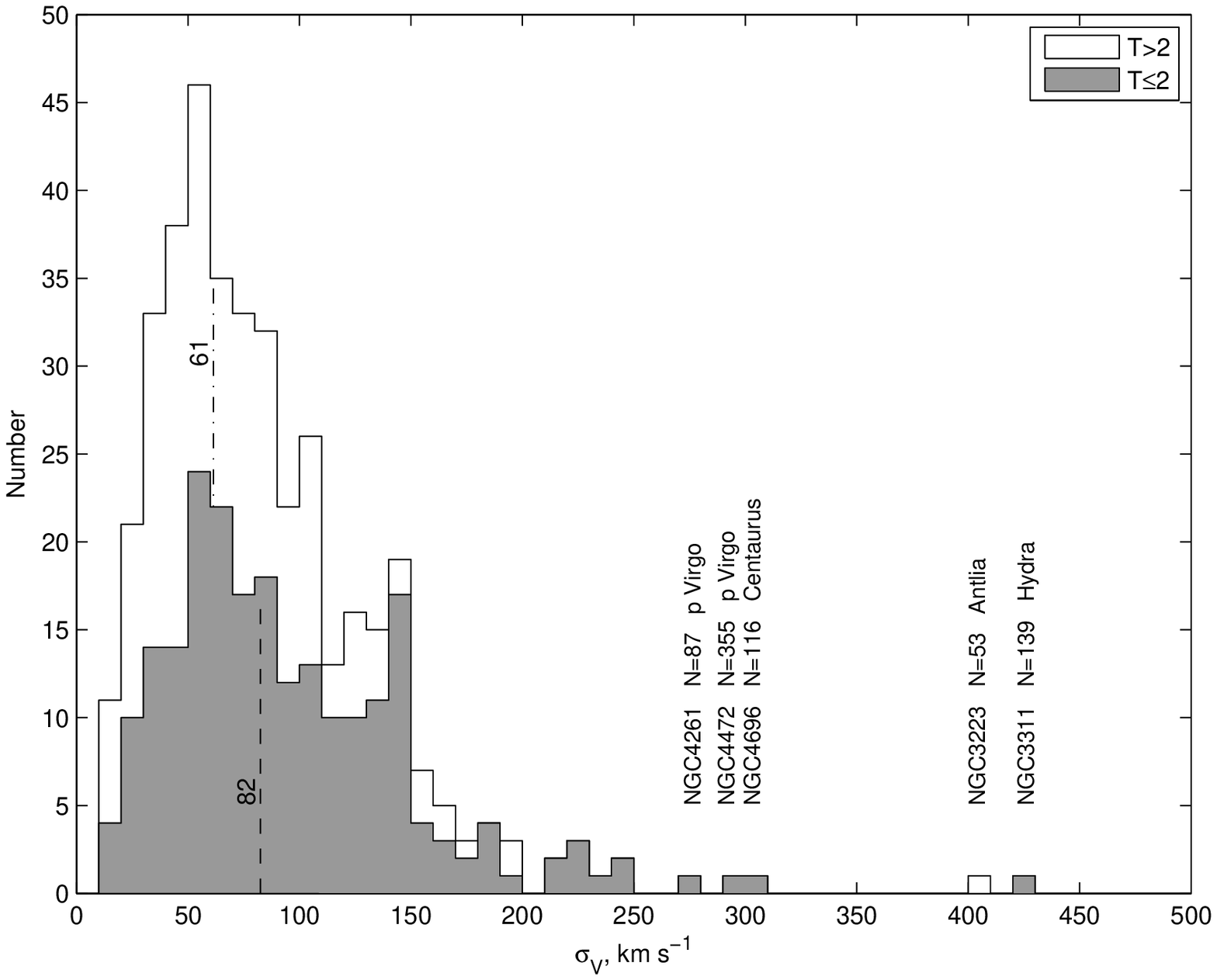}
\caption{
Distribution of the groups by their radial velocity dispersion. 
The groups with a bulge-dominated main galaxy is shaded.
}
\label{fig:h_sigv}
\end{figure*}

Fig.~\ref{fig:h_sigv} shows the distribution of groups by 
their radial velocity dispersion. 
The value $\sigma_V$ in groups ranges from 10 to 450 km s$^{-1}$ 
with the median of 74 km s$^{-1}$. In the groups where
the main galaxy belongs to the E, S0, Sa types, the median
dispersion (82 km s$^{-1}$) is slightly higher than in the groups
with the late-type main galaxy (61 km s$^{-1}$). Since in the
modern optical redshift surveys the typical velocity measurement
errors amount to $\sim 40$ km s$^{-1}$, their effect on the virial
motion amplitude is significant. The groups, presented in the tail
of the distribution are in general those, dominated by the
early-type galaxies; they are marked in the figure in grey.

\begin{figure*}
\includegraphics[width=0.8\textwidth]{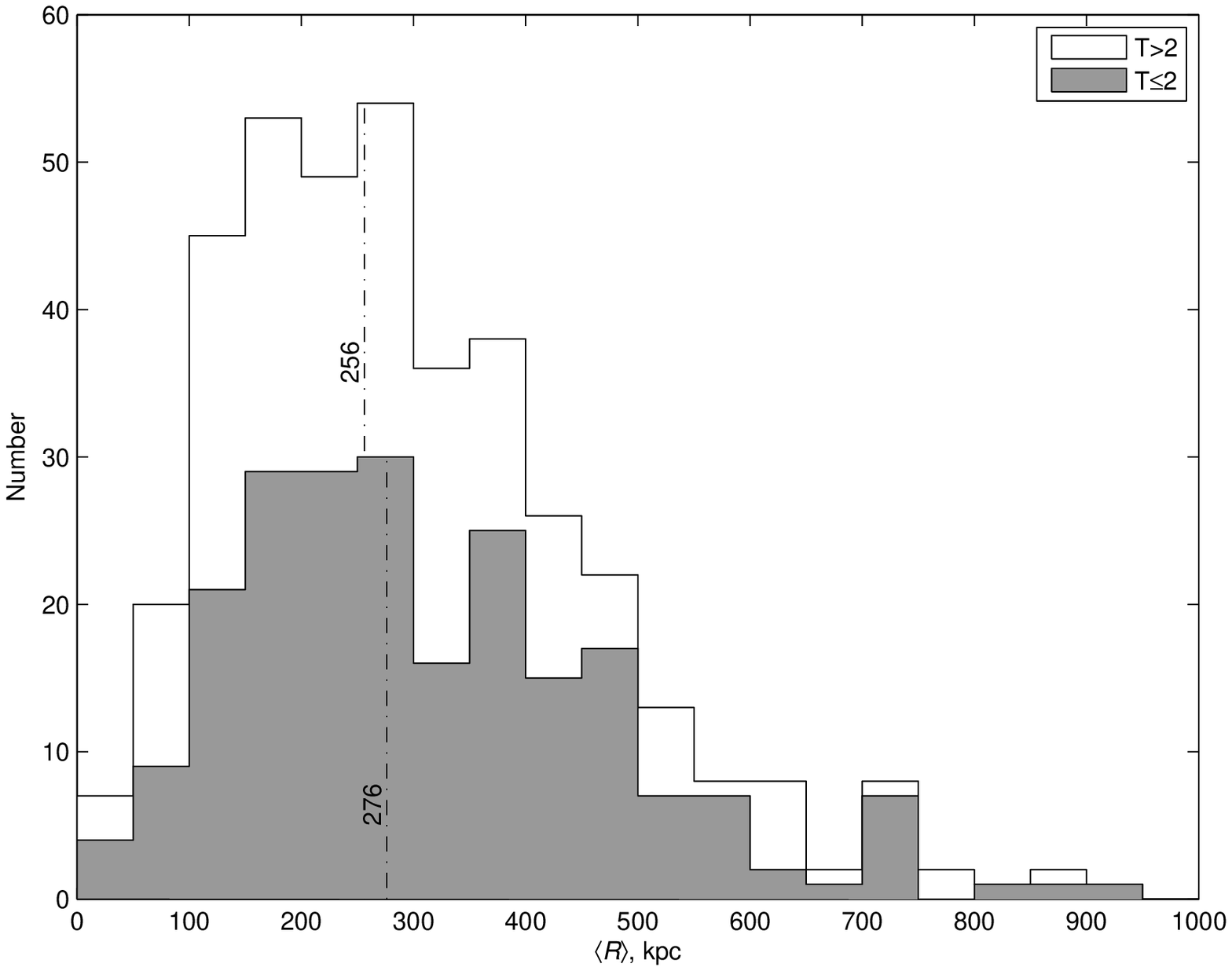}
\caption{
Distribution of the identified groups by their mean projected radius. 
The groups with a bulge-dominated main member are shown in grey.
}
\label{fig:h_r}
\end{figure*}

According to data from Fig.~\ref{fig:h_r}, the mean projected radius of the
groups is distributed over a wide range from 33 kpc to 903 kpc with
a median of 268 kpc. The groups with a dominant early-type galaxy
(shown in grey) have the linear size on the average slightly
larger (276 kpc) than the others (256 kpc). It is obvious that the observed
diversity of linear sizes of groups has a physical origin, 
rather than being caused simply by the projection factors.

\begin{figure*}
\includegraphics[width=0.8\textwidth]{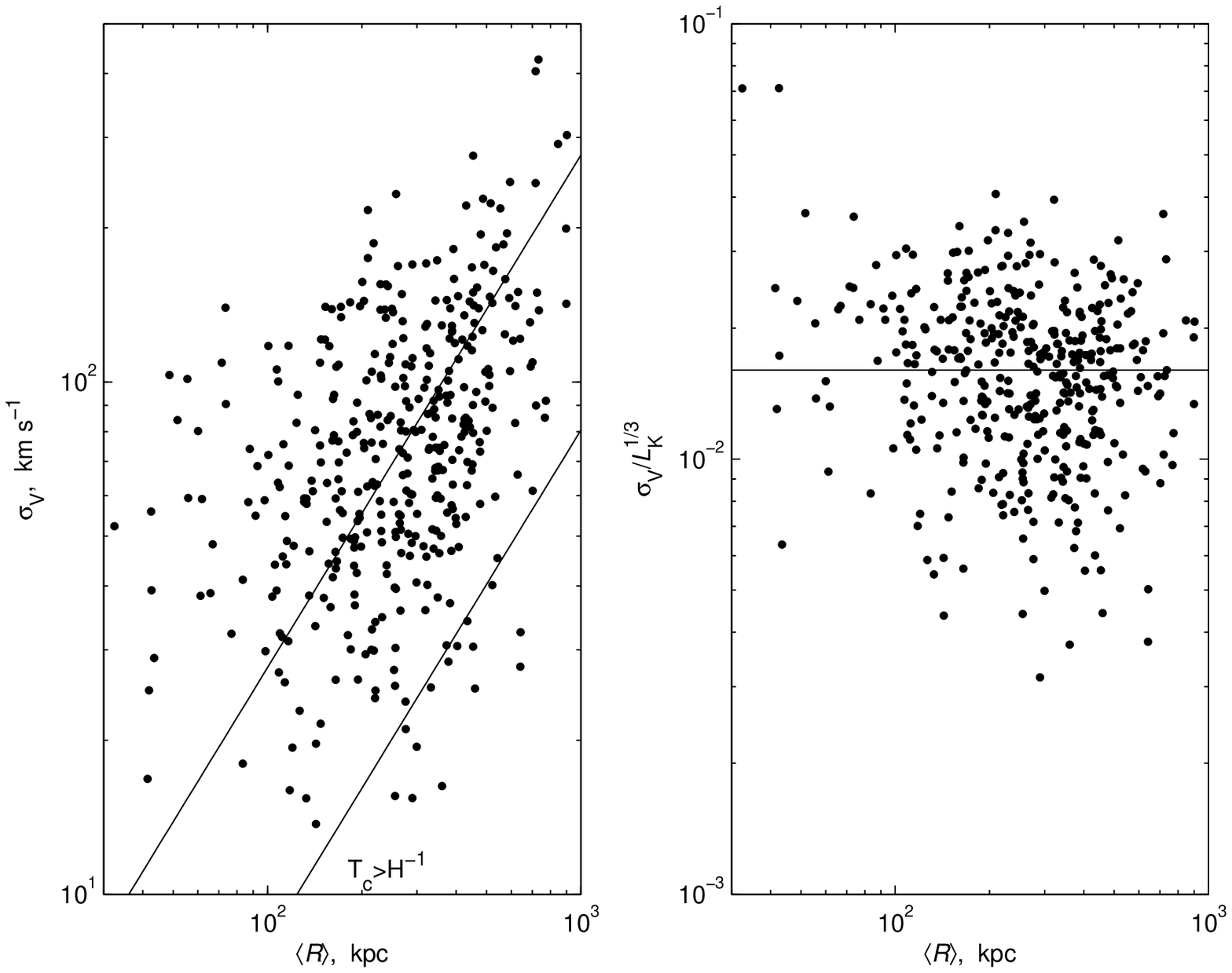}
\caption{
Left: radial velocity dispersion versus mean projected radius for the identified groups.  
Right: the radial velocity dispersion normalized over the total $K$-luminosity of the group.
}
\label{fig:r-sigv}
\end{figure*}

A two-dimensional distribution of 395 groups by linear dimension
and velocity dispersion is presented in the left panel of Fig.~\ref{fig:r-sigv}.
Despite a large scatter, a weak positive correlation between
$\sigma_V$ and $\langle R\rangle$ is visible. 
A straight line in the lower right corner indicates the region 
where the `crossing time' of the group exceeds 
the Hubble time of the Universe $H_0^{-1}$ Gyr. 
Only 3 per cent of all groups outreachs this limit
mainly due to the projection factors. 
The right panel in Fig.~\ref{fig:r-sigv} depicts a similar
distribution, but with the radial velocity dispersion normalized
over the total $K$-luminosity of the group. The straight line
corresponds to the case where the velocity dispersion is
proportional to the linear size of the group, $\sigma_V\propto \langle R\rangle$, 
and the integral luminosity of the group is proportional to
its volume, $L_K\propto\langle R\rangle^3$.

\begin{figure*}
\includegraphics[width=0.8\textwidth]{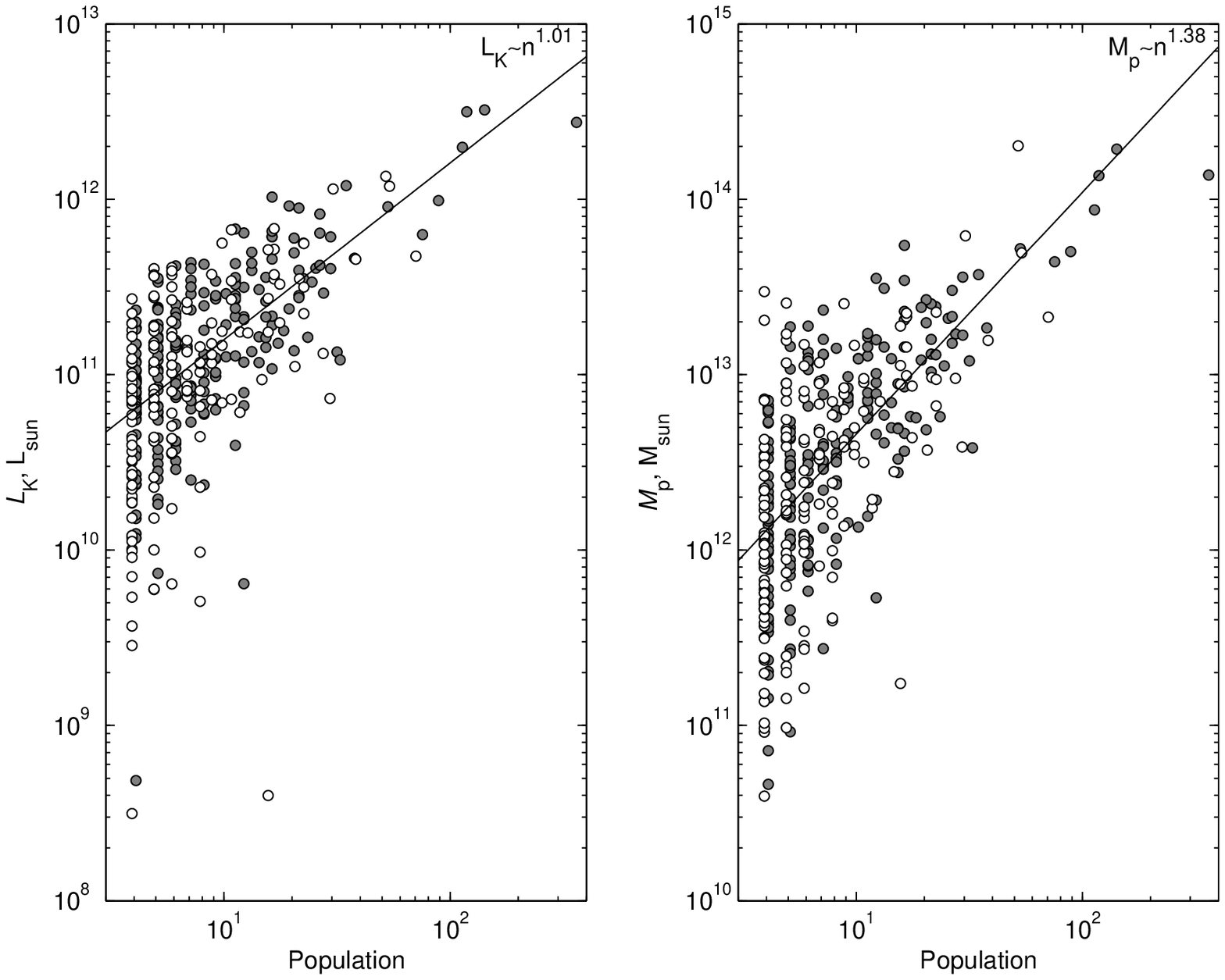}
\caption{
Total $K$-band luminosity (left panel) and projected mass (right panel) 
versus the group population with known velocities. 
The groups with a bulge-dominated principal member are given in grey.
}
\label{fig:n-lum}
\end{figure*}

Fig.~\ref{fig:n-lum} shows how the total $K$-luminosity (left) and projected mass
(right) change with an increasing number of group members. Solid
and open circles denote, respectively, the systems with a
bulge-dominated and disc-shaped main members. The straight lines
in the panels correspond to the linear regressions: 
$\log L_K\propto 1.01\lg n$ and $\log M_p\propto 1.38\lg n$, different
slopes of which indicate that the mass of the group grows with the
population faster than its luminosity.
It might imply that either the baryon fraction is lower in more massive groups 
or, more plausibly, that the cumulative star-formation efficiency is lower.

\begin{figure*}
\includegraphics[width=0.8\textwidth]{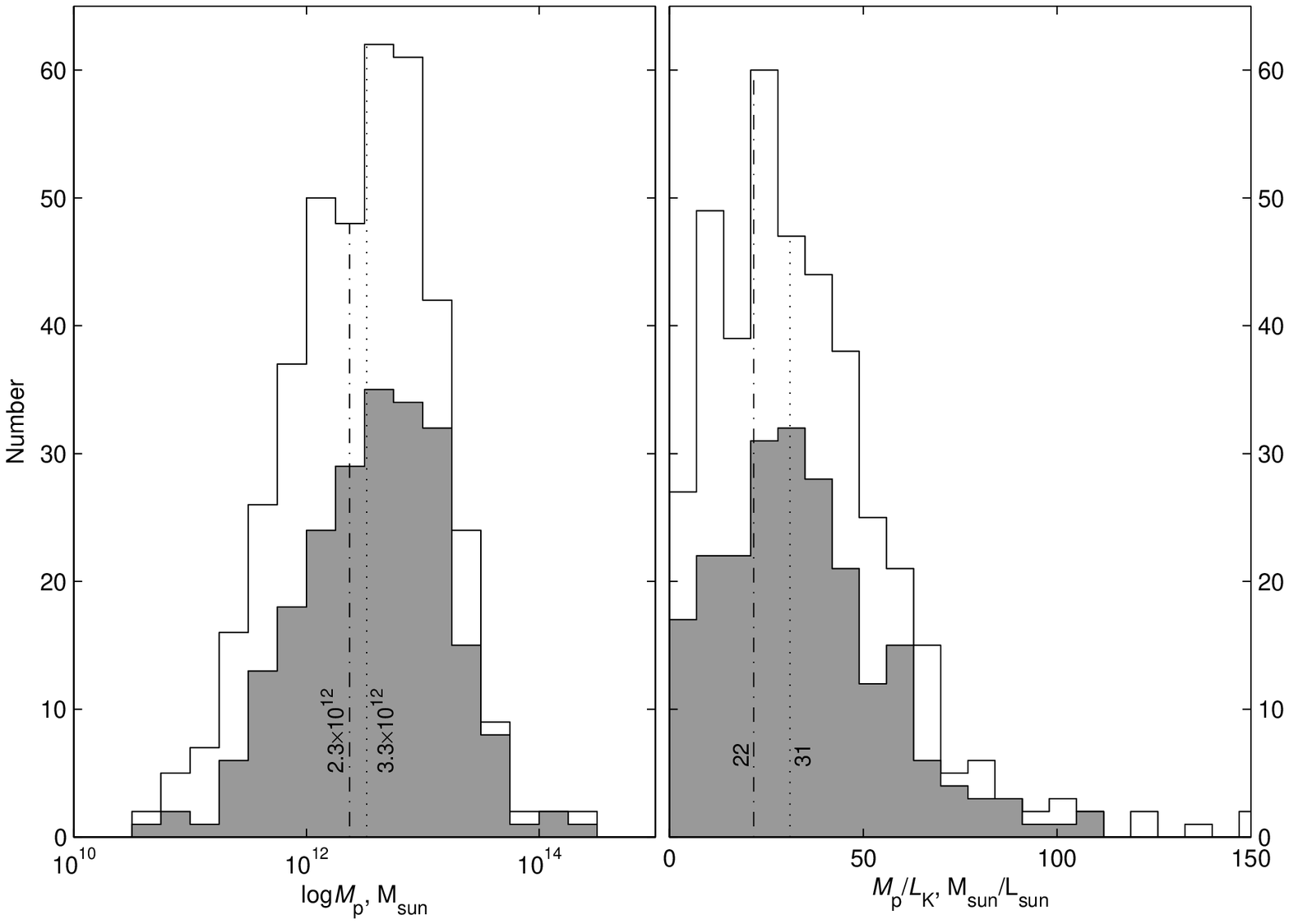}
\caption{
Distribution of the groups by projected mass (left panel) and  projected 
mass-to-K-light ratio (right panel) in the logarithmic scale. 
The dotted vertical line and the dashed one indicate the biased and 
unbiased (corrected for velocity measurement errors) medians, respectively.
}
\label{fig:h_mass}
\end{figure*}

As we have noted, modern optical surveys provide an insufficient
accuracy of radial velocity measurements even for relatively
nearby galaxies. This is reflected in a noticeable way on the
estimates of group masses. A histogram of the distribution of the
number of groups over the logarithm of the projected mass
is demonstrated in the left panel of Fig.~\ref{fig:h_mass}. 
The distribution has a fairly symmetrical shape with the median
$3.3\,10^{12}M_{\odot}$. Here the median mass of the groups
which are dominated by early-type galaxies (the grey part of the
histogram) amounts to $4.0\,10^{12}M_{\odot}$. A transition to
the unbiased mass estimates via a quadratic subtraction from
$\sigma^2_V$ of the velocity measurement errors reduces the median
to $2.3\,10^{12}M_{\odot}$. A similar effect manifests itself
in the distribution of groups by the value of the projected
mass-to-luminosity ratio (right panel in Fig.~\ref{fig:h_mass}), 
where the medians of the biased and unbiased estimates are 31 and 22
$M_{\odot}/L_{\odot}$, respectively. Thus, neglecting the real
accuracy of galaxy velocity measurements, one overestimates the
mass of groups on the average by 30 per cent.
Someone might suspect the discrepancy between adopted value $\kappa=6$, 
based on training the group-finding algorithm in the Local Volume,
and the derived value of $M/L_K=22$ from the groups found with this algorithm.
There is no contradiction here. The criterion (\ref{eq:Criterion1}--\ref{eq:Criterion2})
does not use any assumptions about projection effects. 
These inequalities are true for observable values because projection 
only decreases real separation and velocity difference between galaxies.
For virialized systems the projection factor is $3\pi/2$ for criterion \ref{eq:Criterion1}.
Thus the expected value of $M/L_K$ is about 28 that is in good agreement with
derived value of mass-to-light ratio.

\begin{figure*}
\includegraphics[width=0.8\textwidth]{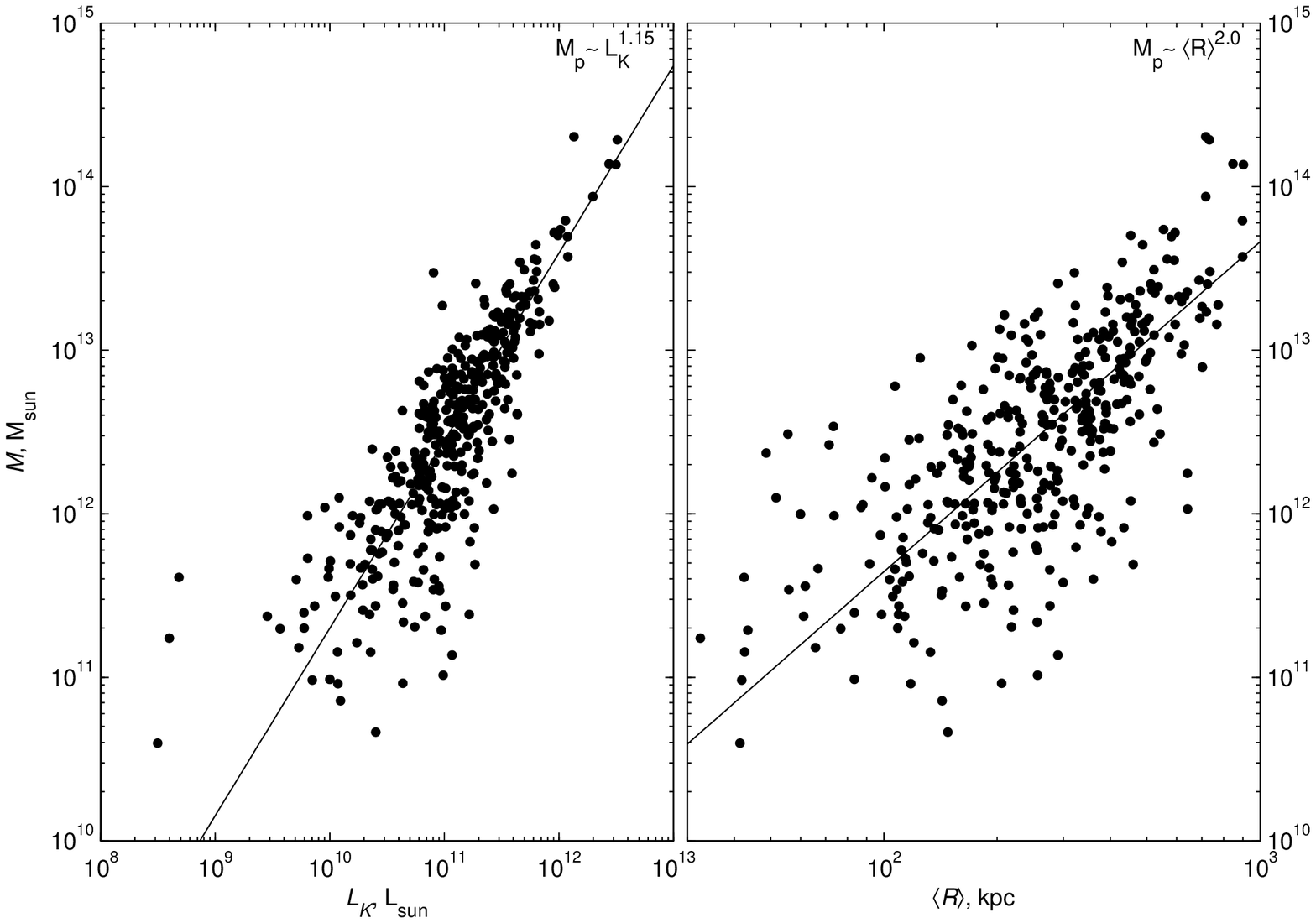}
\caption{
Projected mass of the groups versus their total $K$-band luminosity (left) 
and mean projected radius (right panel). 
The dotted lines show a linear fit to the data.
}
\label{fig:lum-mass}
\end{figure*}

The relations between the projected mass of the group and its
luminosity  $L_K$, as well as its linear size $\langle R\rangle$ are
represented in the panels of Fig.~\ref{fig:lum-mass}.
The lines of \textrm{robust regression weighted by dispersion of scatter} in the
left and right panels are
$\lg M_p\propto 1.15\pm0.03\lg L_K$ and  
$\lg M_p \propto 2.0\pm0.1 \lg \langle R\rangle$, respectively.
Here the correlation of the mass $M_p$ with the linear size of a group
is significantly less pronounced than with the luminosity.

\begin{figure*}
\includegraphics[width=0.8\textwidth]{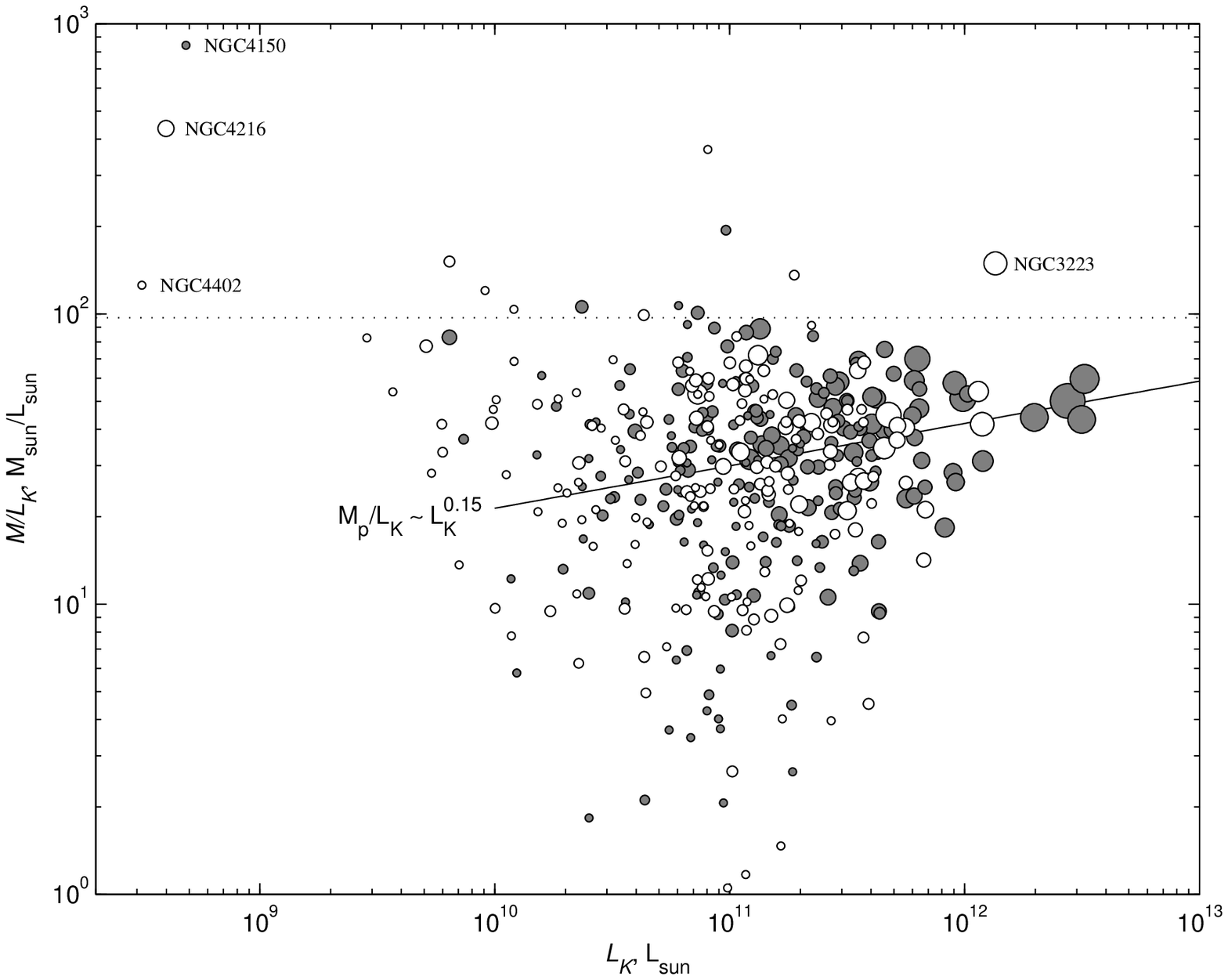}
\caption{
Mass-to-K-light ratio of the groups as a function of $K$-luminosity. 
The groups are represented by circles scaled by population. 
The groups with bulge-dominated main member are shown by filled circles. 
The horizontal dotted line traces the global cosmic ratio, 97 $M_{\odot}/L_{\odot}$, 
corresponding to $\Omega_m = 0.28$. 
The regression (solid) line is drawn taking into account the $K$-luminosity as a statistical weight.
}
\label{fig:lum-ml}
\end{figure*}

Fig.~\ref{fig:lum-ml} demonstrates the distribution of 395 groups 
by the total luminosity $L_K$ and by the  projected mass-to-luminosity ratio. 
The groups, where the dominant galaxy belongs to the early types
$T\leq2$ are shown with solid circles, and the rest of groups
--- with open circles. The size of the circles is proportional to
the group population. The horizontal dotted line captures the
ratio $M/L_K=97$ $M_{\odot}/L_{\odot}$, which corresponds to 
$K$-luminosity density $j_K=4.28\,10^8$ $L_\odot$ Mpc$^{-3}$
\citep{Jones+2006} assuming
Hubble constant $H_0=73$ km s$^{-1}$ Mpc$^{-1}$ 
and
the cosmic parameter of matter density $\Omega_m=0.28$ in the standard
$\lambda$CDM model.

We can see that the region of low values  $M_p/L_K < 5$
$M_{\odot}/L_{\odot}$ is occupied by the groups with the population
$n = 4\textrm{--}6$. An example of this is a group around the spiral
galaxy NGC~660 with $R_h = 235$ kpc and velocity dispersion of
only 29 km s$^{-1}$. In the region of high values $M_p/L_K > 97$
$M_{\odot}/L_{\odot}$ there are 10 groups, most of which with 
$n = 4\textrm{--}6$ members are in this region due to a random projection
factors game. In fact, only one group in the zone $M_p/L_K>97$:
NGC~3223 (Antlia) is a massive group (cluster) with a big number
of measured radial velocities, n=53. In addition, the region of
values $M_p/L_K>97$ contains three groups with low luminosities:
NGC~4216 (n=16), NGC~4402 (n=4) and NGC~4150 (n=4). The first two
groups are located in the Virgo cluster core and have mean radial
velocities of $+55$ km s$^{-1}$ and $+117$ km s$^{-1}$, respectively.
Consistent with these velocities, the distances and luminosity of
these groups are anomalously low, what led to their fictitiously
high $M_p/L_K$ ratios. The disadvantage of our algorithm, where
the distance of a group is determined by the mean radial velocity
of its galaxies, is most pronounced in the regions with large
peculiar motions. Some groups we identified in the Virgo cluster
core are, most probably, false groups, rather than physical
subsystems in the Virgo cluster. The highest value
$M_p/L_K=843 M_{\odot}/L_{\odot}$ falls within the group of 5
galaxies around NGC~4150 in the Coma~I region. The mean radial
velocity of the group, $+211$ km s$^{-1}$, corresponds to the
distance of 2.9 Mpc, whereas the individual distances of NGC~4150
and the other members are 4--5 times more distant. It is likely
that these galaxies with low radial velocities have a large
component of peculiar velocity, moving towards the Virgo cluster.

The line $\lg M_p/L_K \propto 0.15\lg L_K$ 
on Fig.~\ref{fig:lum-ml}
corresponds to correlation between mass and luminosity
from Fig.~\ref{fig:lum-mass}. 
It shows that the mass-luminosity ratio of a group on the
average increases from poor to rich groups. 
Interestingly, its intersection with the $\Omega_m=0.28$ line can 
occur for group with luminosity $L_K \simeq 3\,10^{14}$ $L_{\odot}$, 
which is comparable with total luminosity of the Local Supercluster.

\begin{figure*}
\includegraphics[width=0.8\textwidth]{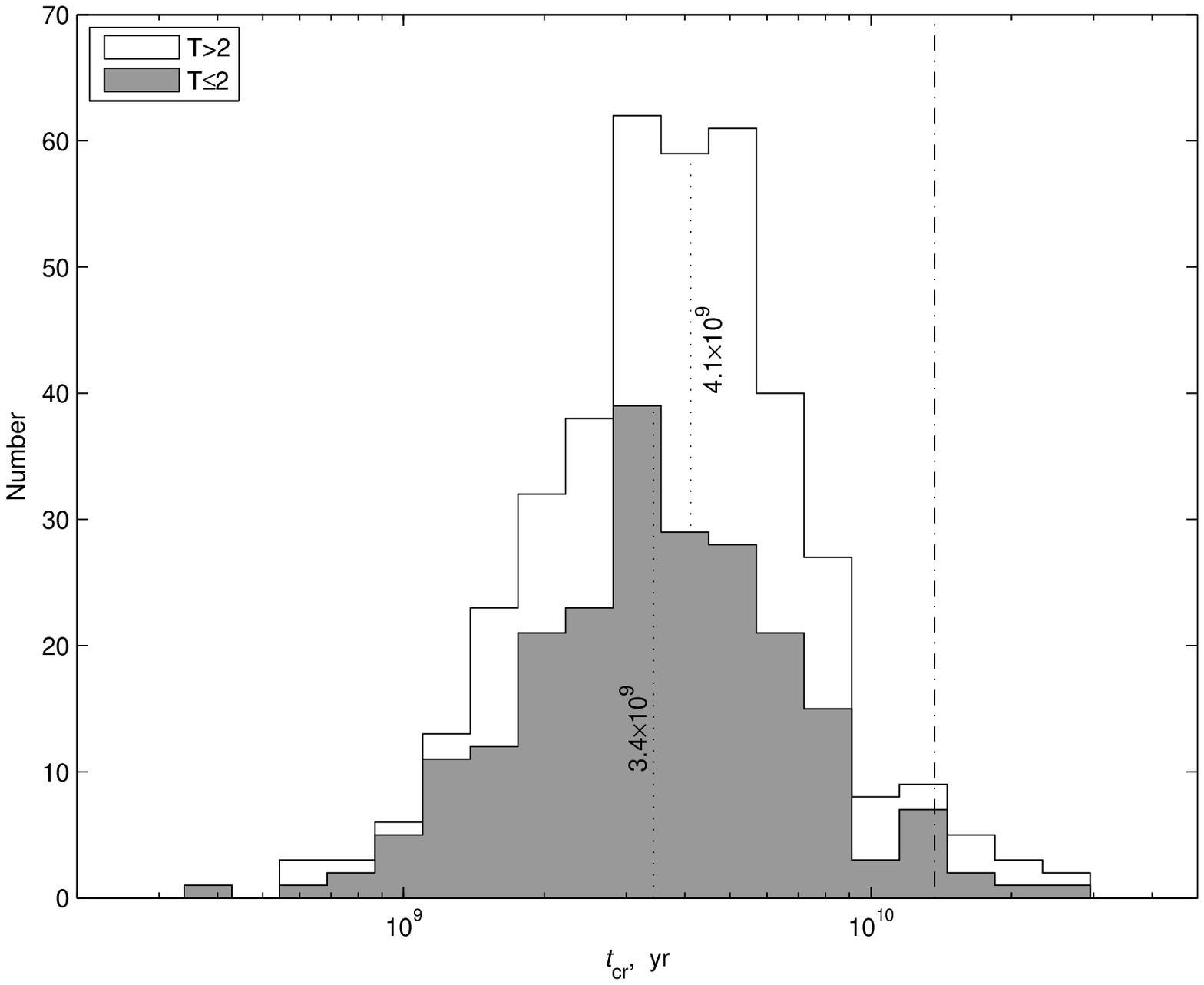}
\caption{
The number of groups with crossing time in the specified intervals. 
The groups with bulge-dominated principal member are shaded. 
The vertical line fixes the age of the Universe, 13.7 Gyr.
}
\label{fig:h_tcr}
\end{figure*}

An important parameter of galaxy system is the crossing time.
Taking into account the effects of projection the crossing time is
\begin{equation}
t_{cr} = \frac{4}{\pi\sqrt{3}}\frac{\langle R_{\bot}\rangle}{\sigma_V}. 
\end{equation}
where the mean projected pairwise separation, $R_{\bot}$, is characteristic size 
and the velocity dispersion, $\sigma_V$, is characteristic inner motion in the group.
The distribution of 395 groups by
$t_{cr}$, presented in Fig.~\ref{fig:h_tcr}, has a fairly symmetrical shape
with the median at 3.8 Gyr. The groups with a dominant E, S0, Sa
galaxy (hatched) are characterized by a slightly shorter crossing
time (3.4 Gyr) than the others (4.1 Gyr). Only 3 per cent of groups fall
into the region $t_{cr}>13.7$ Gyr. 
The correction for velocity error increases this value up to 19 per cent.
Consequently, the essential fraction of the groups 
selected by our criterion can be considered as dynamically
evolved systems.

\section{Discussion}

\begin{figure*}
\includegraphics[width=0.8\textwidth]{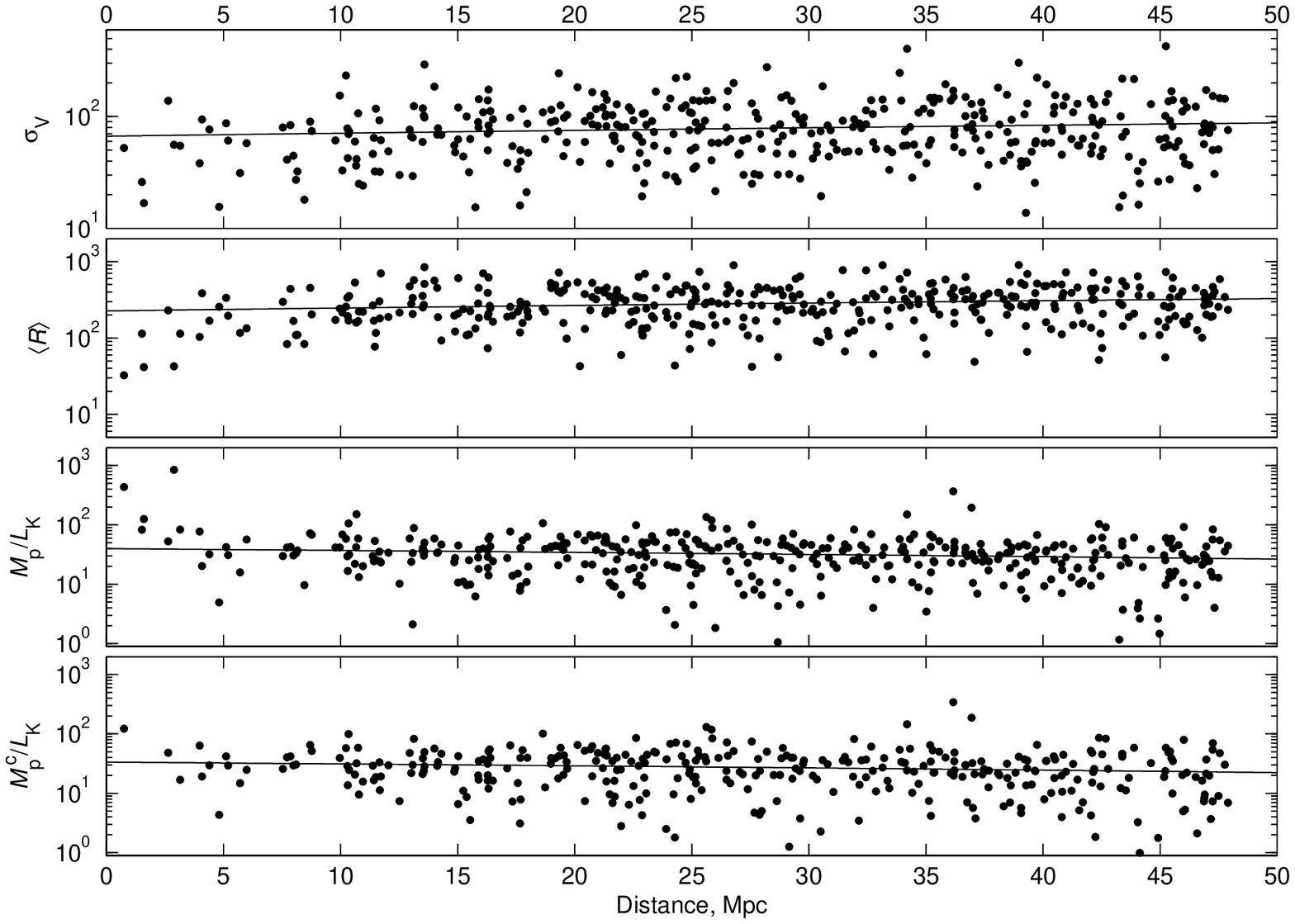}
\caption{
From top to bottom: velocity dispersion, mean projected radius, projected mass-to-light ratio 
and corrected mass-to-light ratio as a function of distance. 
The solid lines show a linear fit to the data.
}
\label{fig:dist}
\end{figure*}

As the data in Fig.~\ref{fig:dist} show, the radial velocity dispersion and the
linear diameter of group trend to increase slightly with distance.
However, the mean projected mass-to-luminosity ratio 
(corrected or uncorrected for velocity measurement errors)
practically does not depend on the distance. In other words,
the physical requirements we used for galaxy clustering (the
negative value of the total energy of a virtual pair, and its
component separation to be within the radius of the zero-velocity
surface) appear to be weakly sensitive to a loss of dwarf galaxies, 
which increases with a distance. 
Therefore, our criterion does not need special tuning of
dimensionless parameter $\kappa$ as a function of the group distance.

It should be remind that the population of groups $n$, presented
in our catalogue, corresponds to the number of group members with
already measured radial velocities. Besides these galaxies, the
group volume may contain a lot of dwarf galaxies without radial
velocity estimates. Such dwarf systems (dSph and dIrr), usually
having a low surface brightness, were found in large quantities in
nearby ($D\leq15$ Mpc) groups by \citet{KK1998,KK2004,KKH2007} at the  visual
examination of the POSS-II prints. Recently, \citet{EigenthalerZeilinger2010}
used the SDSS survey data to search for dwarf
members of the NGC~5846 group. 
\citet{Mahdavi+2005,TT2008,TT2009} looked for the dwarf
population in the groups NGC~5846, NGC~5353/4 and NGC~1023, using
high-resolution images taken with the MegaCam CCD camera on the
CFHT telescope. A large number of dwarf members in the Virgo and
Fornax clusters were found by 
\citet{BST1985,Ferguson1989,FergusonSandage1989,Mieske+2007}. The
physical membership of these objects needs to be confirmed via the
distance and/or radial velocity measurements. A consolidated list
of the most probable dwarf galaxies in the nearby groups from our
catalogue as potential targets for measuring new radial velocities
is under preparation by \citet{Karachentseva+2011}.

A comparison of galaxy membership in our groups with its
affiliation in groups in other catalogues is quite difficult to
perform due to essential differences in the volumes of space
covered by different catalogues, as well as the sets of original
data on radial velocities of galaxies used, which are rapidly
growing with time. Nevertheless, well-known groups, such as
NGC~5371, NGC~5846, show a fairly detailed agreement in their
member composition.

\begin{figure*}
\includegraphics[width=0.8\textwidth]{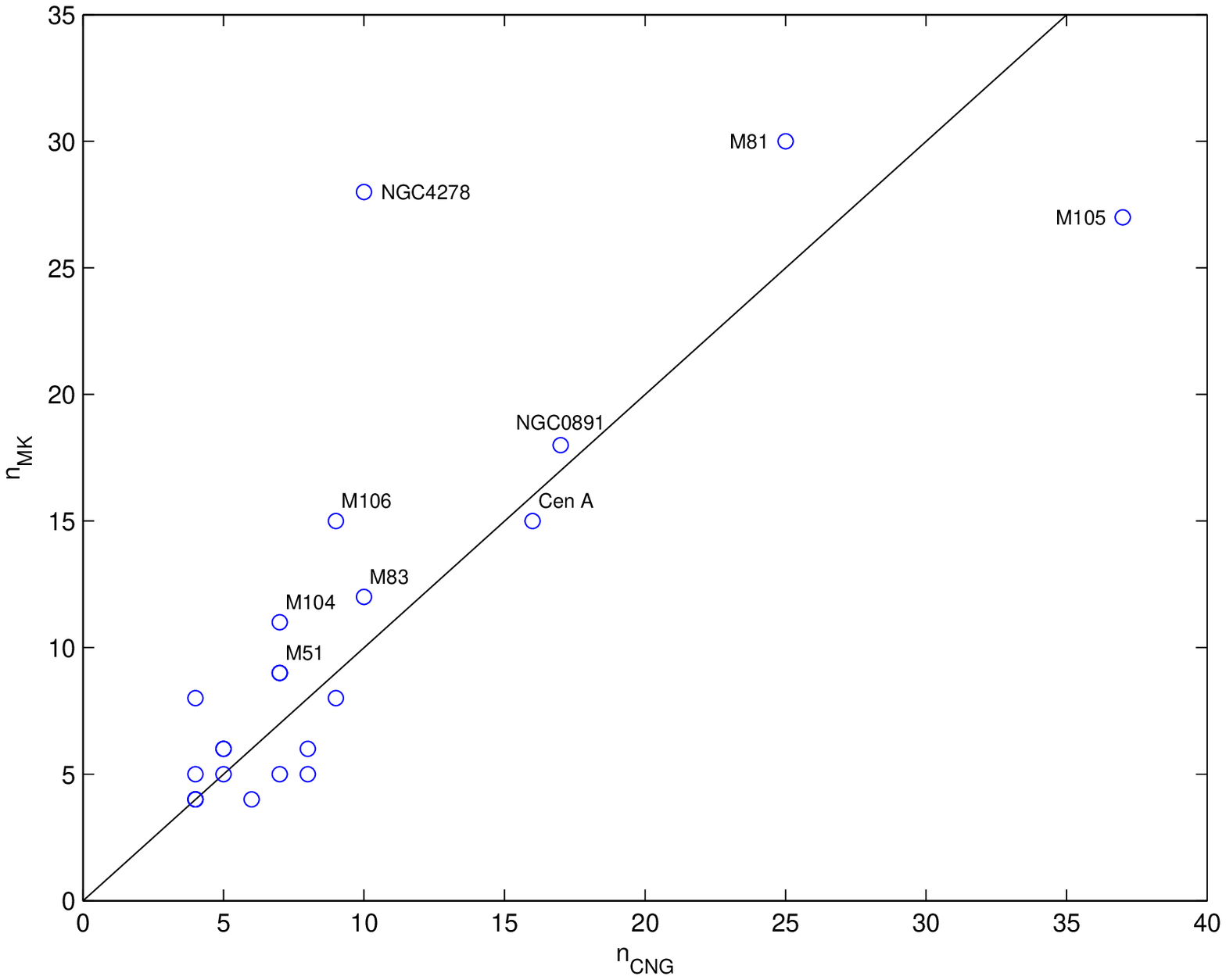}
\caption{
Number of galaxies in the Local Volume groups (with known velocities) 
identified by the clustering algorithm versus their number 
in the same groups as presented in CNG-catalogue \citep{CNG}.
}
\label{fig:cng-mk}
\end{figure*}

Using the data of `A Catalog of Neighboring Galaxies' (CNG) \citep{CNG},
which presents individual
distances to many galaxies with $D<10$ Mpc, and indicates in the
comments their membership in the nearby groups, we compared the
populations of these groups in the old ($n_{CNG}$) and new
$(n_{MK}$) catalogues. As we can see from Fig.~\ref{fig:cng-mk}, for the majority
of 23 nearby groups there is a good agreement between the numbers
of group members. An exception is a group of galaxies NGC~4278
(Coma~I), which is located in a complex region near Virgo, where
several groups overlap at the equator of the Local Supercluster
and a significant role is played by the Virgocentric infall
effect. It should be emphasized, however, that a comparison we
made generally refers to a rather `cold' region of the Local
Volume. An application of our algorithm to the cluster zones
(Virgo, Fornax) with large non-Hubble velocities can generate
fictitious, phantom groups.

\begin{figure*}
\includegraphics[height=0.25\textheight]{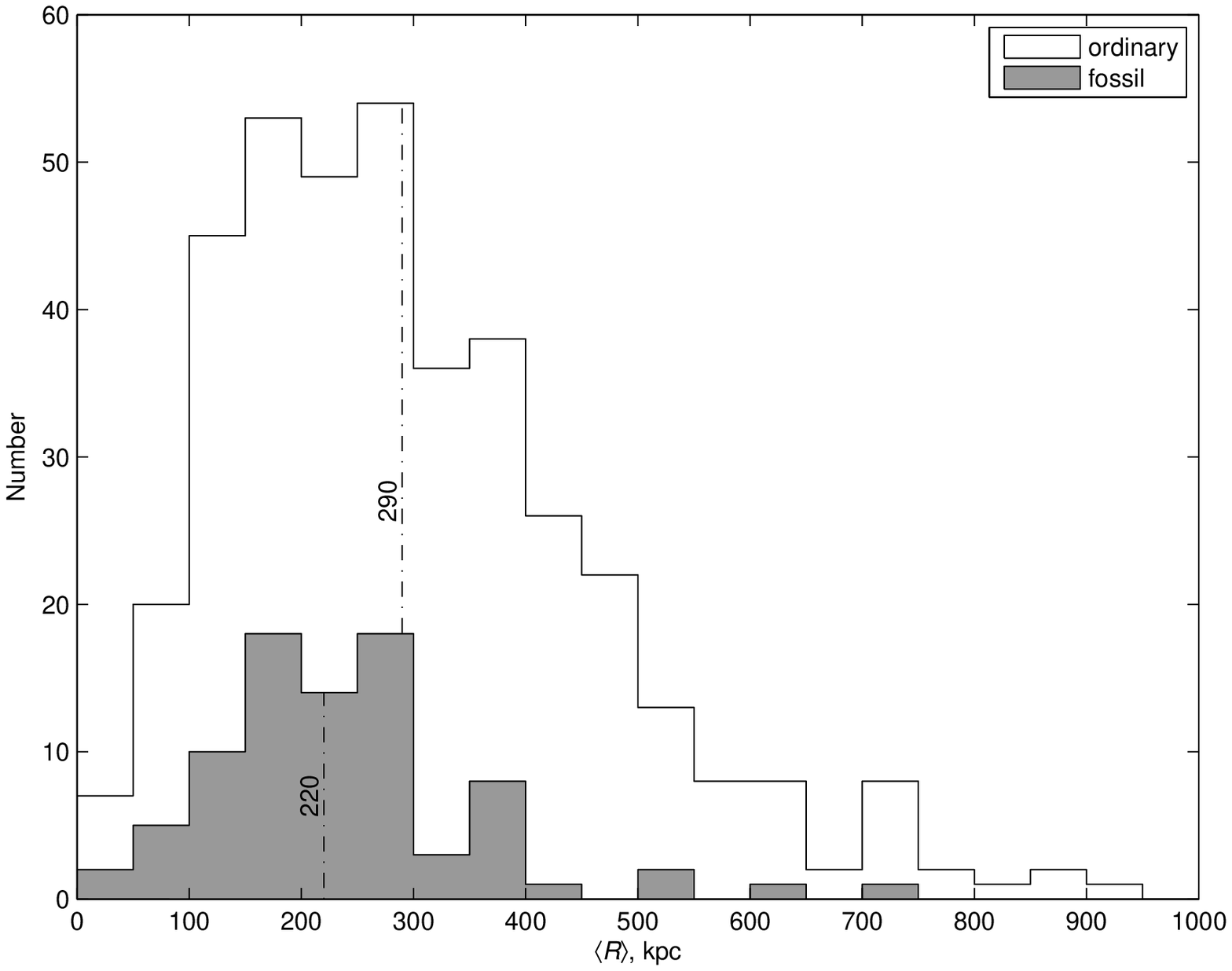}
\includegraphics[height=0.25\textheight]{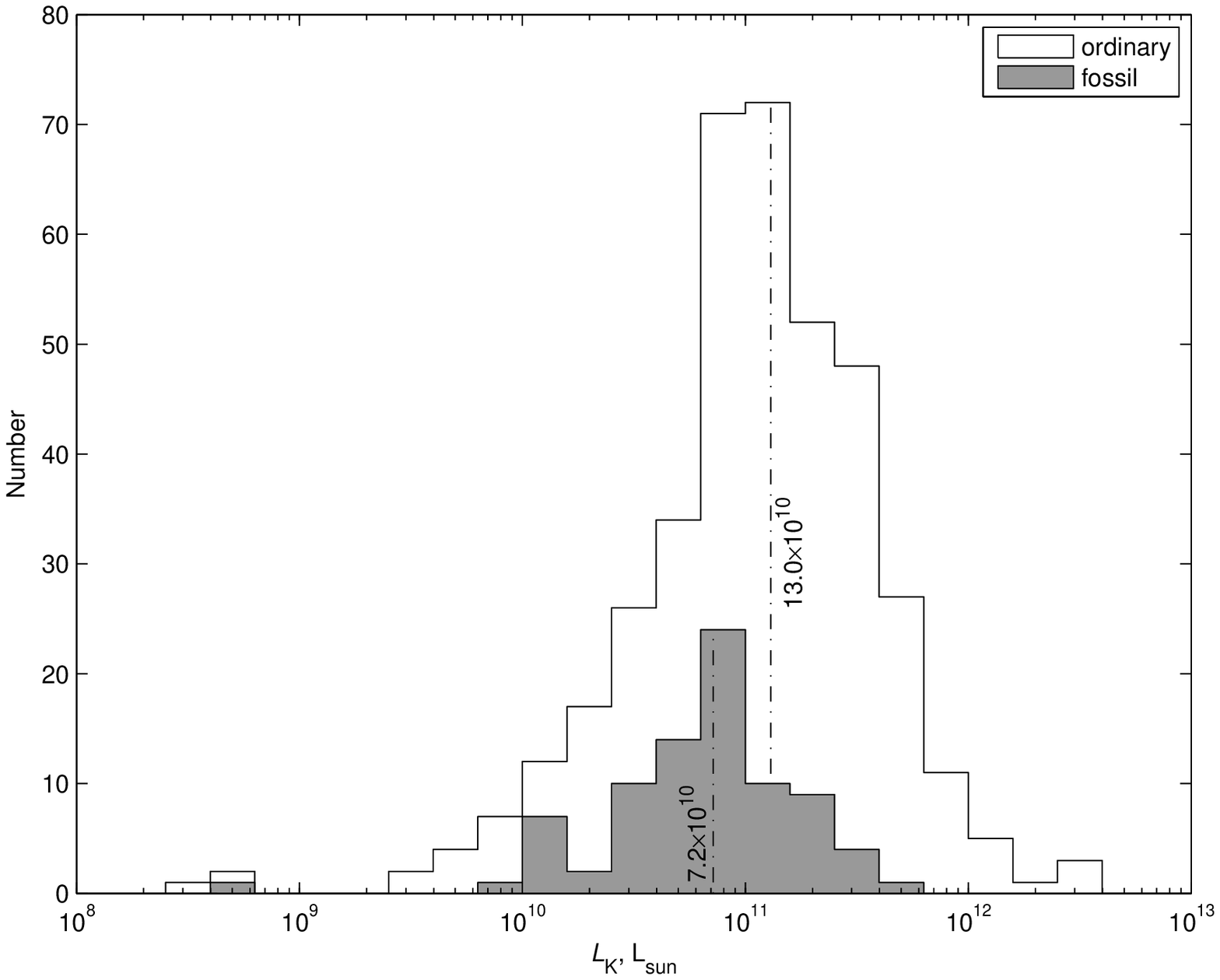}
\includegraphics[height=0.25\textheight]{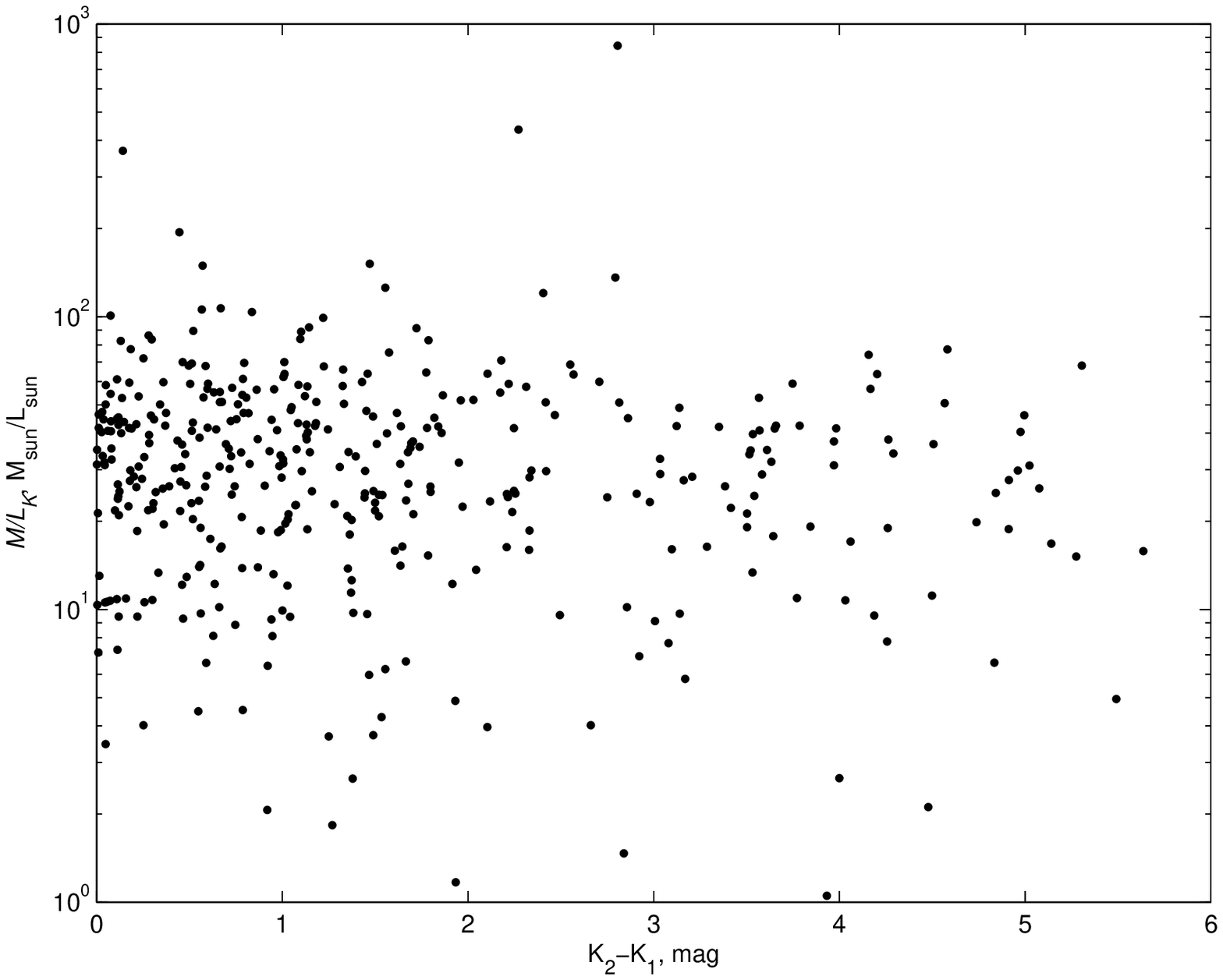}
\caption{
Properties of `fossil' groups (shaded) with respect to others. 
Top left: number of groups versus mean projected radius. 
Top right:  number of groups versus $K$-luminosity. 
Bottom:  projected mass-to-luminosity ratio as a function of K-magnitude 
difference between the first and the second ranked members.
}
\label{fig:fossil}
\end{figure*}

Some authors \citep{Jones+2003,DiazGimenez+2008,vonBendaBeckmann+2008}
propose to distinguish a special
category of `fossil' groups, where the main galaxy significantly
surpasses all the other members on the luminosity scale. The
dynamical evolution of such groups could have evolved on a
different scenario than that of other groups. An example of a
`fossil' group in the Local Volume can be a group of companions
around M~104 (Sombrero), where the K-magnitude difference between
the second and the first brightest galaxies is $K_2-K_1=2.98^m$,
or M~101 with the difference of $K_2-K_1=3.97^m$. The NGC~2903
group has the biggest difference in the Local Volume,
$K_2-K_1=5.64^m$. The groups around M~51 ($K_2-K_1=0.12^m$),
NGC~4244 $(0.18^m)$, NGC~55 $(0.13^m)$, Cen~A $(0.52^m)$ and M~81
$(0.81^m)$ are located at the opposite end of the $K_2-K_1$ scale
in the Local Volume. If we attribute as fossil groups those with
$K_2-K_1>2.50^m$, then our catalogue contains 81 such groups or 21 per cent
of their total number. Two upper panels in Fig.~\ref{fig:fossil} show the
distribution of 395 groups by linear size $\langle R\rangle$ and integral
luminosity $L_K$. Here the fossil groups are given in grey. The
third panel exhibits a relation between the projected
mass-to-luminosity ratio and $K_2-K_1$. The fossil groups with
$K_2-K_1>2.50^m$ tend to have a bit lower mean size, luminosity
and $M_p/L_K$ ratio than the others.

As noted above, our algorithm of galaxy clustering is not entirely
reliable in the `hot' areas with large non-Hubble motions. In
the Virgo as in the core of the Local Supercluster our
criterion identifies 46 groups and pseudogroups. At a lower
density contrast with the parameter $\kappa = 40$ they all merge
into the association M~49, which is easily tracked in column (12)
Table~\ref{tab:groups}. In general this association (i.e., the Virgo cluster)
contains 1558 galaxies with a total luminosity of
$L_K ({\rm Virgo})=1.4\,10^{13}$ $L_{\odot}$ and the projected mass sum of
$\Sigma M_p ({\rm Virgo}) = 6.0\,10^{14}$ $M_{\odot}$. This mass is
close to the total virial mass of Virgo as a single dynamic
aggregate $M_{VIR}=(7\pm1)\,10^{14}M_{\odot}$ according to
\citet{Hoffman+1980,TS1984,Tonry+2000}. 
Note as well that for the conglomerate of the groups
identified by us in the Virgo cluster, the ratio 
$\Sigma M_p/\Sigma L_K=43$ $M_{\odot}/L_{\odot}$ 
practically coincides with the virial estimate 
$M_{VIR}(< 1.8{\rm Mpc})/L_K=(48\pm6)$ $M_{\odot}/L_{\odot}$, obtained by 
\citet{McLaughlin1999} within the radius 1.8 Mpc.

In the region of another nearby cluster, Fornax, our algorithm
identifies 27 multiple systems with the total number of members equal to
379, which at the low density contrast ($\kappa = 40$) join into
the association NGC~1399=Fornax (see the last column of Table~\ref{tab:groups}).
The total luminosity of the Fornax+Eridanus association is 
$\Sigma L_K ({\rm For+Eri}) =5.0\,10^{12}$ $L_{\odot}$, and the sum of projected
masses of 27 systems is 
$\Sigma M_p ({\rm For+Eri}) =2.1\,10^{14}$ $M_{\odot}$. The ratio of these quantities, 
$\Sigma M_p/\Sigma L_K=43$ $M{\odot}/L_{\odot}$ is somewhat higher than the
ratio $32$ $M{\odot}/L_{\odot}$, obtained by \citet{Desai+2004} for
the Fornax cluster itself at 
$L_K ({\rm Fornax}) = 1.8\,10^{12}L_{\odot}$ and 
$M_p ({\rm Fornax}) = 5.9\,10^{13}M_{\odot}$.

We assume that among the 27 multiple systems identified by us in the
Fornax+Eridanus region, the majority are real substructures of the
complex that have not yet reached dynamic equilibrium.
Interestingly, the group NGC~1386 with the mean radial velocity
$+755$ km$s^{-1}$ has the greatest value in the complex,
$M_p/L_K=106$ $M{\odot}/L_{\odot}$. The corresponding Hubble distance
to the group is 10.3 Mpc. However, \citet{Tonry+2001} determined
the individual distances to three galaxies in this group:
NGC~1386, NGC~1395 and E~358--59, and the mean distance deduced
from them is $(19.3\pm1.6)$ Mpc.  Obviously this group is falling
from the far side onto the Fornax cluster. With a more precise
distance, the projected mass-to-luminosity ratio of NGC~1386 group
drops to 57 $M_{\odot}/L_{\odot}$.

A comparison between two rich complexes in the Virgo and
Fornax+Eridanus yielded the ratio of the galaxy number in them as 4.1, 
the ratio of the total K-luminosities as 2.7, 
and the ratio of projected masses as 2.8. 
As a curiosity, note that 
\citet{Crook+2007} estimated the ratio of the projected masses as $M_p
\textrm{(Virgo)}/M_p\textrm{(For+Eri)} = 0.23$, what is contrary to numerous
observational data.

As the most massive structure within the distance of $D=40$ Mpc,
the Virgo cluster amounts to 18 per cent of galaxies in this volume, 
15 per cent of the total $K$-luminosity and 
15 per cent of the projected mass. 
Such a proportion of baryons as well as dark matter contained in rich clusters
is quite consistent with the generally accepted ideas.

In Table~\ref{tab:groups} the group of 9 galaxies around M~51 stands out among
the remaining 394 groups due to its weak isolation. The clustering
of galaxies into associations at a low density contrast results in
a sharp increase in the number of galaxies, joining this group.
Being a group of middle compactness with $R_h=182$ kpc, and
uniting with other groups: NGC~4244, NGC~4258, NGC~4490, NGC~4736,
the M~51 group turns at $\kappa = 40$ into an extended association
with 405 members, the total $K$-luminosity of  $2.0\,10^{12}$ and
the ratio of the projected mass sum to the sum of group luminosities
$\Sigma M_p/\Sigma L_K = 38$ $M_{\odot}/L_{\odot}$. This loose
formation became known in the literature as a Cloud of Galaxies in
the Canes Venatici, CVn~I.

\section{Budget of cosmic matter in the Local universe}

The data on galaxies in groups, pairs and in the field
available at our disposal allow making judgements on the principal
features of the distribution of  bright and dark matter in the
Local universe with the diameter of 80--90 Mpc. 
On this scale we expect to reach the mean density of the Universe with 
accuracy about 15 per cent \citep{PS2010}.

\begin{figure*}
\caption{
Distribution of matter projected onto the supergalactic planes {SGX, SGY} and {SGY, SGZ}. 
The distance scale is given by concentric rings around the Local Group with a step of 10 Mpc.
}
\label{fig:xyz}
\end{figure*}

Fig.~\ref{fig:xyz} reproduces the distribution of matter in the Local
Supercluster and its environs, projected onto the planes \{SGX,SGY\} and 
\{SGY,SGZ\} in the Cartesian supergalactic coordinates.
The distance scale is shown by concentric rings around the Local
Group with a step of 10 Mpc. The horizontal cones in the left
figure are formed by the `Zone of Avoidance' at the galactic
latitudes $|b| < 15^{\circ}$. The Virgo cluster with the
adjacent ridges is located up from the centre at a distance of $Y =17$ Mpc. 
There is a chain of groups/clusters: Antlia, Hydra,
Centaurus seen to the left from Virgo. This structure extends
outside the figure's limits towards the region of the so-called
Great Attractor. The Local Void as a zone of low density is seen
on the upper side (+Z) of the right panel, adjoining with the cone
caused by the Galactic extinction. 
The surface $K$-luminosity density in the units of $M_{\odot}/$Mpc$^2$ is
shown in grey scale.

\begin{figure*}
\includegraphics[width=0.8\textwidth]{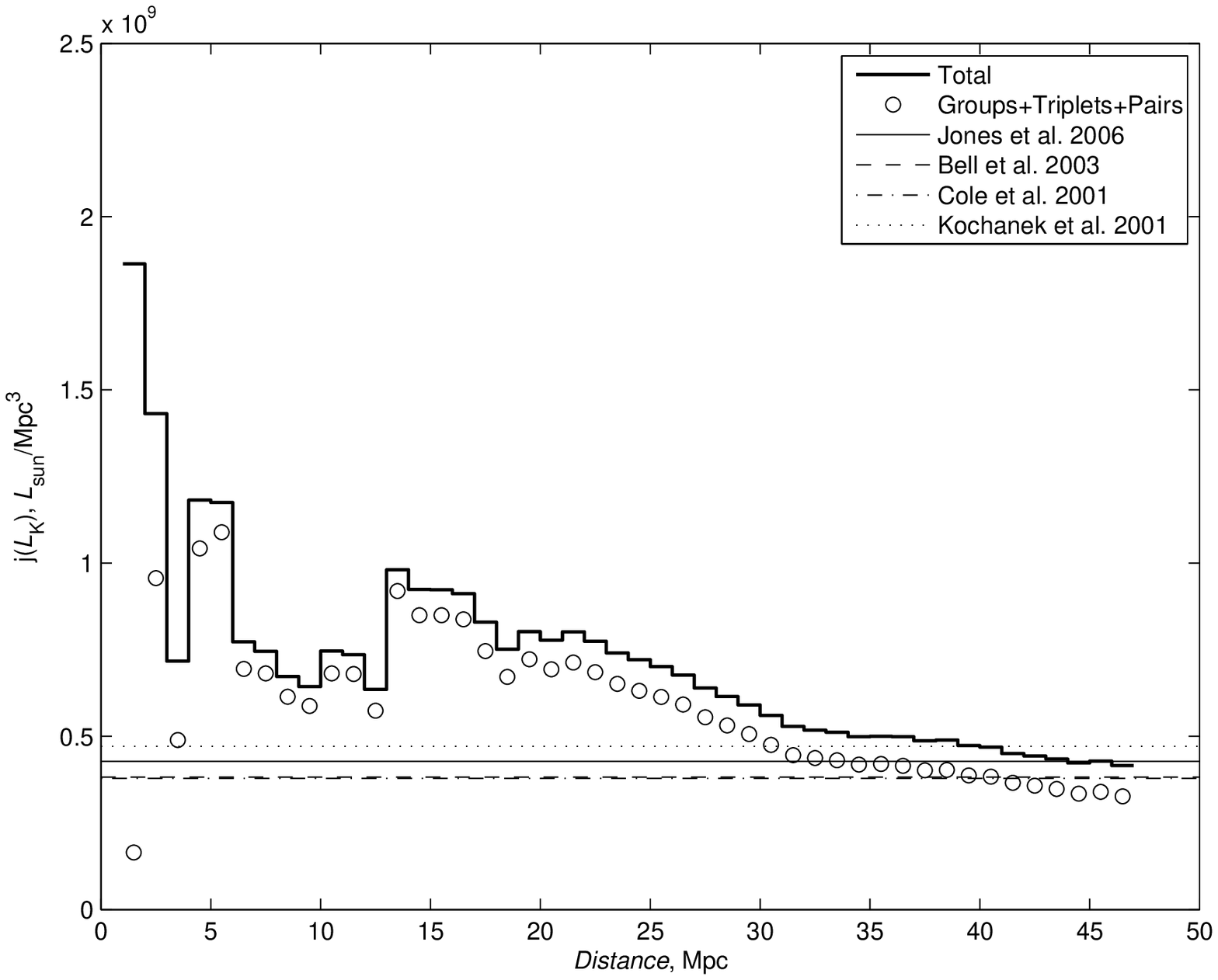}
\caption{
The mean $K$-band luminosity density as a function of distance. 
The density for the groups, triplets and pairs is shown by circles, 
and the total density including field galaxies is plotted by histogram. 
Three horizontal lines represent asymptotic $K$-band luminosity density
from 2MASS by different authors.
}
\label{fig:dist-lum}
\end{figure*}

Fig.~\ref{fig:dist-lum} shows the variation of mean density of $K$-luminosity in
the sphere with the  radius $D_{Mpc}$ and a step of 1 Mpc. The
circles mark the values of the integral luminosity density of the
galaxies belonging to our groups, triplets and pairs. The
histogram reflects the course of the mean density with distance
for all the galaxies, including the galaxies of the field. The
solid, dashed and dotted horizontal lines mark the level of the
mean cosmic $K$-band luminosity density according to 
\citet{Jones+2006,Bell+2003,Cole+2001,Kochanek+2001}, who used
the 2MASS survey data, but with slightly different assumptions on
the proportion of light absorbed by dust inside the galaxies.
Within the 40 Mpc radius sphere, the mean density of $K$-luminosity
according to our data amounts to $4.73\,10^8 L_{\odot}$ Mpc$^{-3}$ 
versus the estimates of the mean space density
$(3.8\textrm{--}4.7)\,10^8$ $L_{\odot}$Mpc$^{-3}$ from the above
sources.

Note that in this volume the total $K$-luminosity of the galaxies
associated into the systems of different multiplicity  $n\geq2$
amounts to 82 per cent of the luminosity of all galaxies. In other words,
only 18 per cent of stellar baryons are located outside the boundaries of
the virial zones of the galaxy systems. 
This is in good accordance with luminosity function measurements 
in groups and overall. For instance, using the data of \citep{Blanton+2003}
on total luminosity density and \citep{ZMM2006} on luminosity function
in groups from SDSS survey and taking into account the fraction of galaxies
in groups (54 per cent), we receive that only 20 per cent of the total light 
is associated with field galaxies.
Comparing the proportion of clustered luminosity in the $K$-band 
(82 per cent) with the relative number of clustered galaxies (54 per cent), 
we conclude that the degree of crowding in dwarf galaxies is 
significantly lower than that in the galaxies with high $K$-luminosity.

Another feature of the Local universe is revealed in that the
approximation of the mean stellar density with increasing $D$ to
its asymptotic cosmic value occurs exactly from the top, i.e. the
broad LG neighbourhood represents a vast zone of overdensity. At
that, the fluctuations
$\delta\rho(L_K)/\langle\rho(L_K)\rangle\sim1$  occur even on the
scales of $\sim(15\textrm{--}20)$ Mpc.

\begin{figure*}
\includegraphics[height=0.28\textheight]{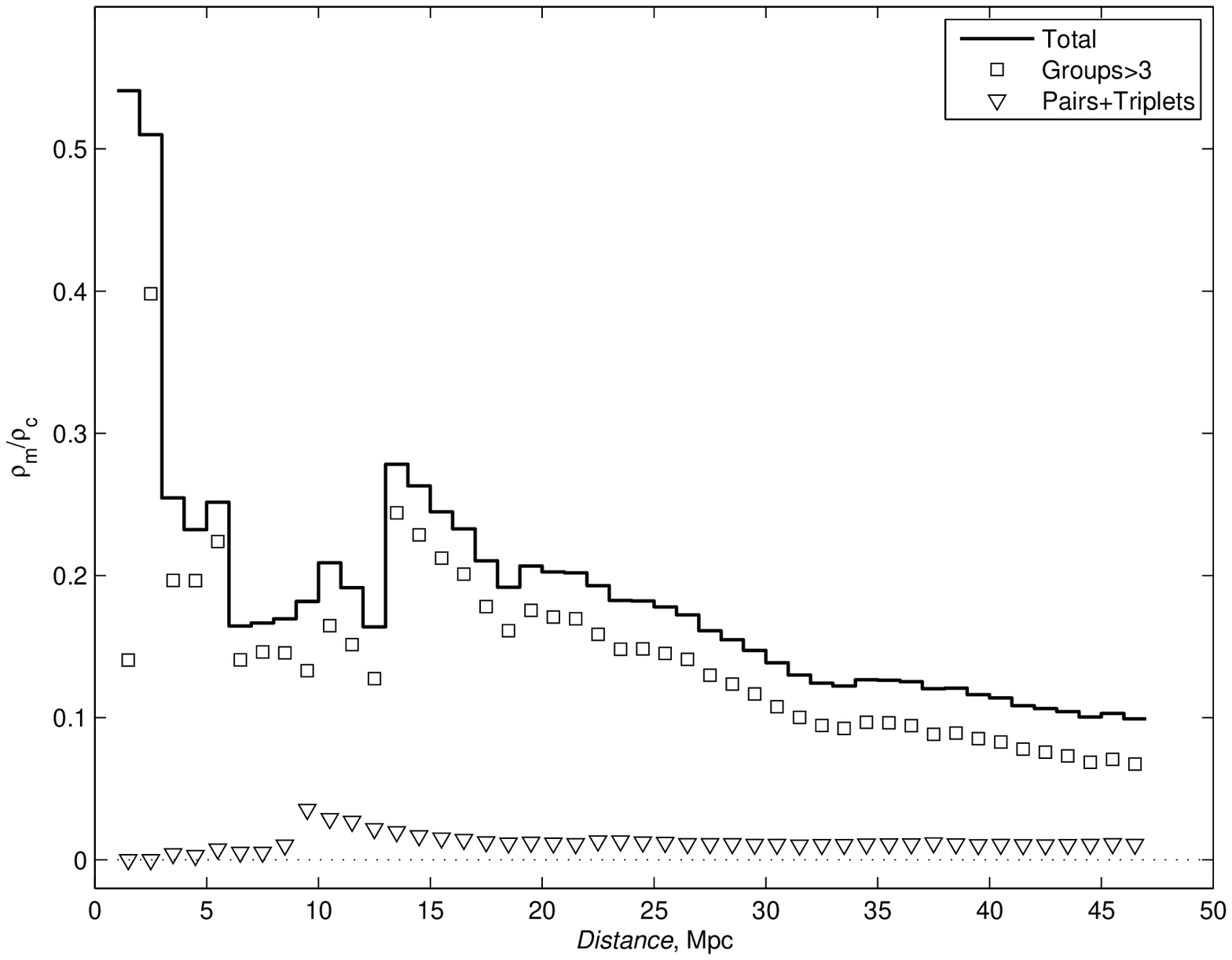}
\includegraphics[height=0.28\textheight]{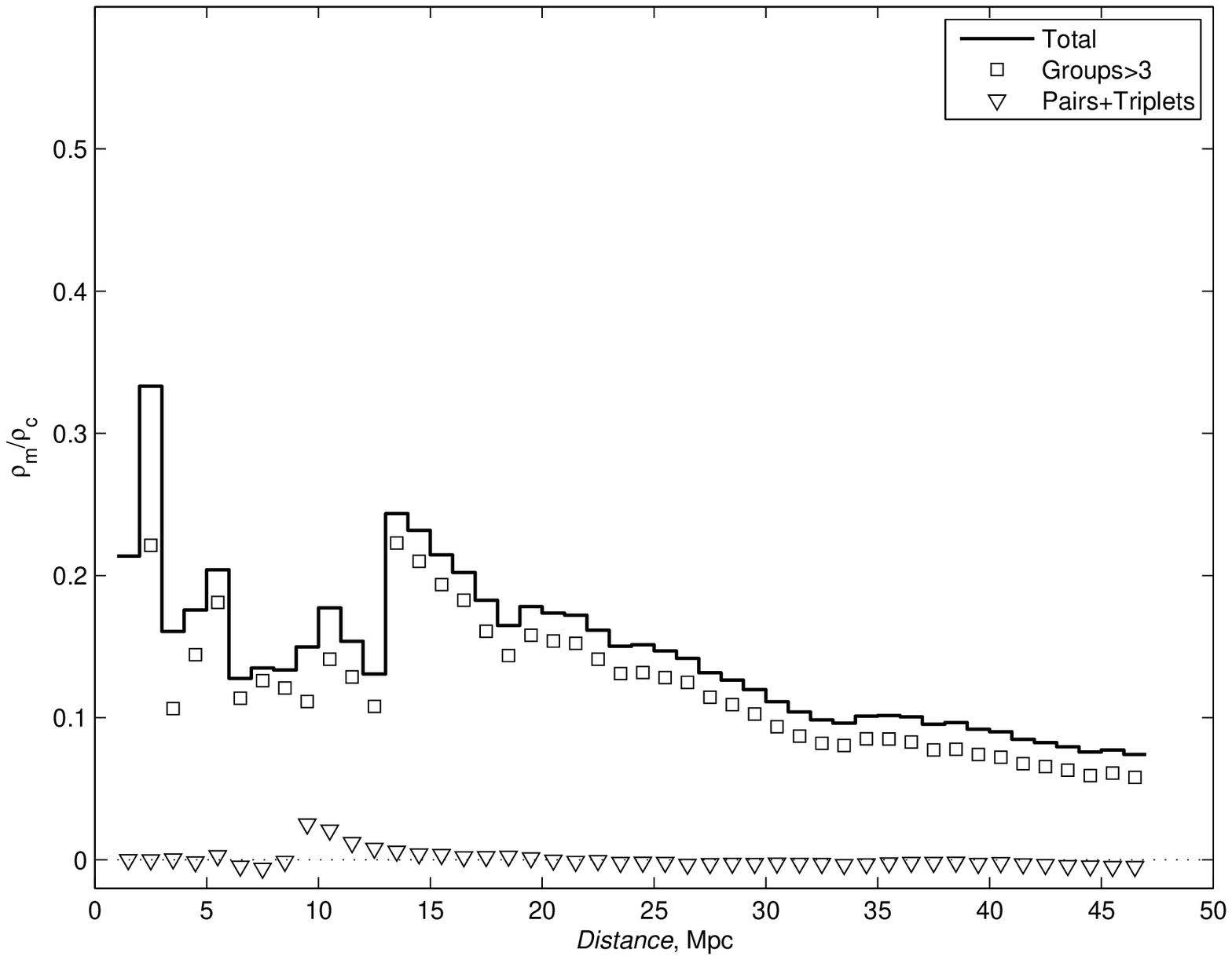}
\caption{
The mean density of matter as a function of distance. 
Contributions of groups and pairs + triplets are shown by squares and triangles, respectively.  
The total mass density is plotted by histogram. 
Left: biased mass estimates ignoring radial velocity errors. 
Right: unbiased matter density corrected for the velocity errors.
}
\label{fig:dist-dens}
\end{figure*}

The distribution of mean density of matter in the Local universe
is shown in two panels of Fig.~\ref{fig:dist-dens} in the units of critical density.
The contribution of the masses produced by pairs and
triplets of galaxies from the catalogues of \citet{Pairs,Triplets}
is shown by the
triangles, and the contribution of the masses of groups
with a population of $n\geq4$ is marked by the squares. Apart from
the group members, we have to take into account the contribution
to the total mass made by the field galaxies. As noted above,
their relative number amounts to 46 per cent, and relative luminosity in
the $K$-band is only 18 per cent. We believe that the isolated galaxies
differ little in their physical properties from the galaxies in
binary or triple systems, as corroborated by the mass estimates
from the effects of weak gravitational lensing 
\citep{Mandelbaum+2006}. 
Attributing to the isolated galaxies the same total
mass-luminosity ratios as those possessed by the components of
low-multiplicity systems, we have depicted the behaviour of the
total matter density in the volumes of radii $D_{Mpc}$ in the form
of a histogram. Compiling the catalogues of groups, triplets and
pairs of galaxies, we were limited by the systems, the mean radial
velocities of which satisfy the condition $\langle
V_{LG}\rangle\leq3500$ km s$^{-1}$, to which at 
$H_0=73$ km s$^{-1}$Mpc$^{-1}$ corresponds the distance of $D$ =47.9 Mpc.
However, in the process of galaxy clustering we took into account
the galaxies beyond this volume with $V_{LG}<4000$ km s$^{-1}$.
Therefore, the values of $\Omega_m$ in the $D=40\textrm{--}47$ Mpc distance
range can be regarded as not highly susceptible to various
boundary effects. The histogram in the top panel of Fig.~\ref{fig:dist-dens} shows
that the mean local density of matter within the distance of
40--47 Mpc amounts to $\Omega_m=0.116\textrm{--}0.096$.

As noted above, the galaxy redshift surveys (CfA, 2dF, 6dF, SDSS)
contain a `noise' component  $\epsilon_V\simeq40$ km s$^{-1}$,
induced by the velocity measurement errors. It affects the 
mass estimates in an asymmetric way, overestimating them on the
average by 30 per cent. The distribution of the mean density of matter,
based on the statistically unbiased estimates of the projected mass
is presented in the bottom panel of Fig.~\ref{fig:dist-dens}. The designations here
are the same as those on the top panel. In the volumes of radii
from 40 to 47 Mpc the mean corrected values of the local density
of matter lie in the range of $\Omega_{m,{\rm loc}}=0.092\textrm{--}0.073$,
which is 3--4 times lower than the global cosmological density
$\Omega_{m}=0.28\pm0.03$ in the standard $\lambda$CDM model
\citep{FukugitaPeebles2004}.

In addition to the galaxy velocity measurement errors, there is
another source of uncertainty in the estimation of the local value
of $\Omega_{m}$. As we have noted, the clustering algorithm we
used does not work reliably enough in the regions of large
non-Hubble motions. Summing the mass of groups we identified in
the Virgo cluster volume ($6.0\,10^{14}M_{\odot}$), we found
that it was smaller than the virial mass estimate
$7.0\,10^{14}$ $M_{\odot}$. In the case of the Fornax cluster, the
total mass of the groups turned out to be larger than the virial
mass ($2.1\,10^{14}$ $M_{\odot}$ versus $0.6\,10^{14}$ $M_{\odot}$).
Therefore, at the total mass of the local sphere with the radius
of 40 Mpc equal to ($2.7\textrm{--}3.4)\,10^{15}M_{\odot}$, the errors
of mass determination $\sim2\,10^{14}M_{\odot}$ result in the
error in the $\Omega_m$ estimations of less than 10 per cent.

Another reason, distorting the local estimate of $\Omega_m$, can
be the fact that we neglected the galaxies that do not yet have
their radial velocities measured. In order to determine the
significance of their contribution, we assumed that the luminosity
function of galaxies is described by the \citet{Schechter1976} function 
with the parameters $M_K^{*} = -24.3$ and $\alpha = -1.0 $
\citep{Bell+2003,Cole+2001}. In the galaxy redshift surveys 
(6dF, SDSS) the completeness of radial velocity measurements is assured
up to the limiting apparent magnitude $K_s \simeq 12.0^m$.
At the boundary of the volume under consideration (40--47) Mpc,
this corresponds to the absolute magnitude of $(-21.0)$, by  $3^m$
fainter than $M_K^{*}$. With the above parameters of the luminosity
function, the contribution of the fainter galaxies in the
integral luminosity on the edge of the considered volume is only about 12 per cent. 
In general, the correction to the mean luminosity density over the entire volume
is less than 10 per cent. Consequently, the neglect of galaxies without
measured radial velocities, and an imperfection of the group
criterion we used may not be the causes of such a strong
difference in the estimates of local and global densities of
matter.

\section{Concluding remarks}

The main objective of our work was to create a catalogue of galaxy
groups, covering the entire sky and extending up to the distances
of 40--45 Mpc from the Local Group. To this end, we used the
clustering algorithm that takes into account individual
characteristics of galaxies. Applying this criterion to the
observational data on 10914 galaxies with the velocities
$V_{LG}<3500$ km s$^{-1}$, we have identified 509 pairs, 168
triple systems and 395 groups with the populations of four or more
members. Totally these samples include 54 per cent of all galaxies, or
82 per cent of the total luminosity of the Local universe of the given
volume. At that, we have not used any special conditions that
would restrict the initial sample of galaxies by morphological
type or other characteristics (apart from the availability of
radial velocity measurements). Minimal selectivity of our sample
makes it attractive for the analysis of various properties of
galaxy systems depending on their environment.

We consider that the most important result here is the estimation
of the mean density of matter in the Local universe. According to
our data, its value amounts to   $\Omega_m=0.08\pm0.02$ in the
spherical volume with the diameter of 80--90 Mpc. As noted above,
the low estimates of  $\Omega_m\sim 0.10$ also follow from the
virial masses in galaxy groups derived from the lists by 
\citet{Vennik1984,NGC,Magtesyan1988}. In a recent paper by
\citet{Crook+2007} the estimates of $\Omega_m$ were obtained
ranging from 0.10 to 0.23, depending on the choice of parameters
for galaxy association into groups, and on the way of computing
the virial masses of groups. Earlier, \citet{Bahcall+2000}
studied the published estimates of the mass-luminosity ratio in
the groups and clusters, and deduced the value
$\Omega_m=0.16\pm0.05$. Recently, \citet{AbateErdogdu2009} analysed
the peculiar velocity field for the 2MASS galaxies, using the
SFI++ survey data \citep{Springob+2007}, and concluded that in
the Local universe with a characteristic scale of $\sim6000$ km s$^{-1}$ 
the density of matter lies within the interval of $\Omega_m=0.09\textrm{--}0.23$.

At least, three different ideas can be proposed to explain the drastic
difference between the local and the global estimates of $\Omega_m$.
\begin{enumerate}
\item
Dark matter in groups and clusters extends far beyond their virial
radius traced by galaxies. To reduce the $\Omega_m$ discrepancy, 
one ought to assume that the total mass of each group and cluster is 
about 3 times its virial mass. 
However, as it was shown by \citet{K2005,KN2010,Nasonova+2011}, 
the total mass of nearby groups and Virgo, Fornax clusters within the radius
of zero velocity surface $R_0$ is almost the same as their virial mass.
Note that $R_0$ is $\sim (3.5\textrm{--}4.0)R_\textrm{vir}$. 
Therefore, the existence of large amount of dark matter at periphery of 
systems is inconsistent with the observational data.
\item
One can image a possibility that the local Universe is
significantly (3 times) under-dense relative to the global density.
The largest structure in the Universe detected to date is the Sloan
Great Wall, filamentary aggregation of galaxies of 450 Mpc long, 
at distance of 310 Mpc \citep{Gott+2005}. 
In principle, we can reside in a ``valley'' between such kind large scale walls. 
But, Fig.~\ref{fig:dist-lum} tells us that we are living in the luminous matter 
overdensity extending till to about 40 Mpc. 
Of course, this local overdensity can be only a peak surrounded by wider under-dense region. 
However, numerous $K$-band counts of galaxies in the range of $K = 12\textrm{--}19^m$ 
\citep{Djorgovski+1995,BLK1998,Huang+2001,Totani+2001} do not show
the presence of any significant cosmic lacuna around us within $\sim2000$ Mpc.
\item
It is possible that essential part of dark matter in the Universe
(about 2/3) is scattered outside virial (and even collapsing) regions
being concentrated in dark clumps or distributed as a homogeneous
dark ``sea''. Some observational arguments favourable to the existence
of massive dark clumps have been presented by \citet{NS2004,Jee+2005} from weak lensing, 
and by \citet{KKH2006} from properties of disturbed isolated galaxies.
\end{enumerate}

\section*{Acknowledgments}

The authors are grateful to V.E. Karachentseva for the great work on the verification and 
refinement of the original observational data.
We are thankful to Brent Tully, Stefan Gottl{\"o}ber and Helene Courtois for discussions and useful comments.
We acknowledge the usage of the HyperLEDA database\footnote{\url{http://leda.univ-lyon1.fr}}. 
This research has made use of the NASA/IPAC Extragalactic Database\footnote{\url{http://nedwww.ipac.caltech.edu/}}
which is operated by the Jet Propulsion Laboratory, California Institute of Technology, 
under contract with the National Aeronautics and Space Administration.
We acknowledge the usage of the HIPASS public data release\footnote{\url{http://www.atnf.csiro.au/research/multibeam/release/}}.
This work was supported by the RFBR grants 10--02--00123, 08--02--00627, RUS-UKR 09--02--90414.

\bibliographystyle{mn2e}
\bibliography{grps}   

\begin{thebibliography}{81}
\expandafter\ifx\csname natexlab\endcsname\relax\def\natexlab#1{#1}\fi

\bibitem[{{Abate} \& {Erdo{\u g}du}(2009)}]{AbateErdogdu2009}
{Abate} A., {Erdo{\u g}du} P., 2009, \mnras, 400, 1541

\bibitem[{{Abazajian} {et~al.}(2009){Abazajian}, {Adelman-McCarthy},
  {Ag{\"u}eros}, {Allam}, {Allende Prieto}, {An}, {Anderson}, {Anderson},
  {Annis}, {Bahcall}, {Bailer-Jones}, {Barentine}, {Bassett}, {Becker},
  {Beers}, {Bell}, {Belokurov}, {Berlind}, {Berman}, {Bernardi}, {Bickerton},
  {Bizyaev}, {Blakeslee}, {Blanton}, {Bochanski}, {Boroski}, {Brewington},
  {Brinchmann}, {Brinkmann}, {Brunner}, {Budav{\'a}ri}, {Carey}, {Carliles},
  {Carr}, {Castander}, {Cinabro}, {Connolly}, {Csabai}, {Cunha}, {Czarapata},
  {Davenport}, {de Haas}, {Dilday}, {Doi}, {Eisenstein}, {Evans}, {Evans},
  {Fan}, {Friedman}, {Frieman}, {Fukugita}, {G{\"a}nsicke}, {Gates},
  {Gillespie}, {Gilmore}, {Gonzalez}, {Gonzalez}, {Grebel}, {Gunn},
  {Gy{\"o}ry}, {Hall}, {Harding}, {Harris}, {Harvanek}, {Hawley}, {Hayes},
  {Heckman}, {Hendry}, {Hennessy}, {Hindsley}, {Hoblitt}, {Hogan}, {Hogg},
  {Holtzman}, {Hyde}, {Ichikawa}, {Ichikawa}, {Im}, {Ivezi{\'c}}, {Jester},
  {Jiang}, {Johnson}, {Jorgensen}, {Juri{\'c}}, {Kent}, {Kessler}, {Kleinman},
  {Knapp}, {Konishi}, {Kron}, {Krzesinski}, {Kuropatkin}, {Lampeitl},
  {Lebedeva}, {Lee}, {Lee}, {Leger}, {L{\'e}pine}, {Li}, {Lima}, {Lin}, {Long},
  {Loomis}, {Loveday}, {Lupton}, {Magnier}, {Malanushenko}, {Malanushenko},
  {Mandelbaum}, {Margon}, {Marriner}, {Mart{\'{\i}}nez-Delgado}, {Matsubara},
  {McGehee}, {McKay}, {Meiksin}, {Morrison}, {Mullally}, {Munn}, {Murphy},
  {Nash}, {Nebot}, {Neilsen}, {Newberg}, {Newman}, {Nichol}, {Nicinski},
  {Nieto-Santisteban}, {Nitta}, {Okamura}, {Oravetz}, {Ostriker}, {Owen},
  {Padmanabhan}, {Pan}, {Park}, {Pauls}, {Peoples}, {Percival}, {Pier}, {Pope},
  {Pourbaix}, {Price}, {Purger}, {Quinn}, {Raddick}, {Fiorentin}, {Richards},
  {Richmond}, {Riess}, {Rix}, {Rockosi}, {Sako}, {Schlegel}, {Schneider},
  {Scholz}, {Schreiber}, {Schwope}, {Seljak}, {Sesar}, {Sheldon}, {Shimasaku},
  {Sibley}, {Simmons}, {Sivarani}, {Smith}, {Smith}, {Smol{\v c}i{\'c}},
  {Snedden}, {Stebbins}, {Steinmetz}, {Stoughton}, {Strauss}, {Subba Rao},
  {Suto}, {Szalay}, {Szapudi}, {Szkody}, {Tanaka}, {Tegmark}, {Teodoro},
  {Thakar}, {Tremonti}, {Tucker}, {Uomoto}, {Vanden Berk}, {Vandenberg},
  {Vidrih}, {Vogeley}, {Voges}, {Vogt}, {Wadadekar}, {Watters}, {Weinberg},
  {West}, {White}, {Wilhite}, {Wonders}, {Yanny}, {Yocum}, {York}, {Zehavi},
  {Zibetti}, \& {Zucker}}]{SDSS7}
{Abazajian} K.~N., {Adelman-McCarthy} J.~K., {Ag{\"u}eros} M.~A., {Allam}
  S.~S., {Allende Prieto} C., {An} D., {Anderson} K.~S.~J., {Anderson} S.~F.,
  {Annis} J., {Bahcall} N.~A., {Bailer-Jones} C.~A.~L., {Barentine} J.~C.,
  {Bassett} B.~A., {Becker} A.~C., {Beers} T.~C., {Bell} E.~F., {Belokurov} V.,
  {Berlind} A.~A., {Berman} E.~F., {Bernardi} M., {Bickerton} S.~J., {Bizyaev}
  D., {Blakeslee} J.~P., {Blanton} M.~R., {Bochanski} J.~J., {Boroski} W.~N.,
  {Brewington} H.~J., {Brinchmann} J., {Brinkmann} J., {Brunner} R.~J.,
  {Budav{\'a}ri} T., {Carey} L.~N., {Carliles} S., {Carr} M.~A., {Castander}
  F.~J., {Cinabro} D., {Connolly} A.~J., {Csabai} I., {Cunha} C.~E.,
  {Czarapata} P.~C., {Davenport} J.~R.~A., {de Haas} E., {Dilday} B., {Doi} M.,
  {Eisenstein} D.~J., {Evans} M.~L., {Evans} N.~W., {Fan} X., {Friedman} S.~D.,
  {Frieman} J.~A., {Fukugita} M., {G{\"a}nsicke} B.~T., {Gates} E., {Gillespie}
  B., {Gilmore} G., {Gonzalez} B., {Gonzalez} C.~F., {Grebel} E.~K., {Gunn}
  J.~E., {Gy{\"o}ry} Z., {Hall} P.~B., {Harding} P., {Harris} F.~H., {Harvanek}
  M., {Hawley} S.~L., {Hayes} J.~J.~E., {Heckman} T.~M., {Hendry} J.~S.,
  {Hennessy} G.~S., {Hindsley} R.~B., {Hoblitt} J., {Hogan} C.~J., {Hogg}
  D.~W., {Holtzman} J.~A., {Hyde} J.~B., {Ichikawa} S., {Ichikawa} T., {Im} M.,
  {Ivezi{\'c}} {\v Z}., {Jester} S., {Jiang} L., {Johnson} J.~A., {Jorgensen}
  A.~M., {Juri{\'c}} M., {Kent} S.~M., {Kessler} R., {Kleinman} S.~J., {Knapp}
  G.~R., {Konishi} K., {Kron} R.~G., {Krzesinski} J., {Kuropatkin} N.,
  {Lampeitl} H., {Lebedeva} S., {Lee} M.~G., {Lee} Y.~S., {Leger} R.~F.,
  {L{\'e}pine} S., {Li} N., {Lima} M., {Lin} H., {Long} D.~C., {Loomis} C.~P.,
  {Loveday} J., {Lupton} R.~H., {Magnier} E., {Malanushenko} O., {Malanushenko}
  V., {Mandelbaum} R., {Margon} B., {Marriner} J.~P., {Mart{\'{\i}}nez-Delgado}
  D., {Matsubara} T., {McGehee} P.~M., {McKay} T.~A., {Meiksin} A., {Morrison}
  H.~L., {Mullally} F., {Munn} J.~A., {Murphy} T., {Nash} T., {Nebot} A.,
  {Neilsen} E.~H., {Newberg} H.~J., {Newman} P.~R., {Nichol} R.~C., {Nicinski}
  T., {Nieto-Santisteban} M., {Nitta} A., {Okamura} S., {Oravetz} D.~J.,
  {Ostriker} J.~P., {Owen} R., {Padmanabhan} N., {Pan} K., {Park} C., {Pauls}
  G., {Peoples} J., {Percival} W.~J., {Pier} J.~R., {Pope} A.~C., {Pourbaix}
  D., {Price} P.~A., {Purger} N., {Quinn} T., {Raddick} M.~J., {Fiorentin}
  P.~R., {Richards} G.~T., {Richmond} M.~W., {Riess} A.~G., {Rix} H., {Rockosi}
  C.~M., {Sako} M., {Schlegel} D.~J., {Schneider} D.~P., {Scholz} R.,
  {Schreiber} M.~R., {Schwope} A.~D., {Seljak} U., {Sesar} B., {Sheldon} E.,
  {Shimasaku} K., {Sibley} V.~C., {Simmons} A.~E., {Sivarani} T., {Smith}
  J.~A., {Smith} M.~C., {Smol{\v c}i{\'c}} V., {Snedden} S.~A., {Stebbins} A.,
  {Steinmetz} M., {Stoughton} C., {Strauss} M.~A., {Subba Rao} M., {Suto} Y.,
  {Szalay} A.~S., {Szapudi} I., {Szkody} P., {Tanaka} M., {Tegmark} M.,
  {Teodoro} L.~F.~A., {Thakar} A.~R., {Tremonti} C.~A., {Tucker} D.~L.,
  {Uomoto} A., {Vanden Berk} D.~E., {Vandenberg} J., {Vidrih} S., {Vogeley}
  M.~S., {Voges} W., {Vogt} N.~P., {Wadadekar} Y., {Watters} S., {Weinberg}
  D.~H., {West} A.~A., {White} S.~D.~M., {Wilhite} B.~C., {Wonders} A.~C.,
  {Yanny} B., {Yocum} D.~R., {York} D.~G., {Zehavi} I., {Zibetti} S., {Zucker}
  D.~B., 2009, \apjs, 182, 543

\bibitem[{{Bahcall} {et~al.}(2000){Bahcall}, {Cen}, {Dav{\'e}}, {Ostriker}, \&
  {Yu}}]{Bahcall+2000}
{Bahcall} N.~A., {Cen} R., {Dav{\'e}} R., {Ostriker} J.~P., {Yu} Q., 2000,
  \apj, 541, 1

\bibitem[{{Barnes} {et~al.}(2001){Barnes}, {Staveley-Smith}, {de Blok},
  {Oosterloo}, {Stewart}, {Wright}, {Banks}, {Bhathal}, {Boyce}, {Calabretta},
  {Disney}, {Drinkwater}, {Ekers}, {Freeman}, {Gibson}, {Green}, {Haynes}, {te
  Lintel Hekkert}, {Henning}, {Jerjen}, {Juraszek}, {Kesteven}, {Kilborn},
  {Knezek}, {Koribalski}, {Kraan-Korteweg}, {Malin}, {Marquarding}, {Minchin},
  {Mould}, {Price}, {Putman}, {Ryder}, {Sadler}, {Schr{\"o}der}, {Stootman},
  {Webster}, {Wilson}, \& {Ye}}]{HIPASS}
{Barnes} D.~G., {Staveley-Smith} L., {de Blok} W.~J.~G., {Oosterloo} T.,
  {Stewart} I.~M., {Wright} A.~E., {Banks} G.~D., {Bhathal} R., {Boyce} P.~J.,
  {Calabretta} M.~R., {Disney} M.~J., {Drinkwater} M.~J., {Ekers} R.~D.,
  {Freeman} K.~C., {Gibson} B.~K., {Green} A.~J., {Haynes} R.~F., {te Lintel
  Hekkert} P., {Henning} P.~A., {Jerjen} H., {Juraszek} S., {Kesteven} M.~J.,
  {Kilborn} V.~A., {Knezek} P.~M., {Koribalski} B., {Kraan-Korteweg} R.~C.,
  {Malin} D.~F., {Marquarding} M., {Minchin} R.~F., {Mould} J.~R., {Price}
  R.~M., {Putman} M.~E., {Ryder} S.~D., {Sadler} E.~M., {Schr{\"o}der} A.,
  {Stootman} F., {Webster} R.~L., {Wilson} W.~E., {Ye} T., 2001, \mnras, 322,
  486

\bibitem[{{Bell} {et~al.}(2003){Bell}, {McIntosh}, {Katz}, \&
  {Weinberg}}]{Bell+2003}
{Bell} E.~F., {McIntosh} D.~H., {Katz} N., {Weinberg} M.~D., 2003, \apjs, 149,
  289

\bibitem[{{Bershady} {et~al.}(1998){Bershady}, {Lowenthal}, \& {Koo}}]{BLK1998}
{Bershady} M.~A., {Lowenthal} J.~D., {Koo} D.~C., 1998, \apj, 505, 50

\bibitem[{{Binggeli} {et~al.}(1985){Binggeli}, {Sandage}, \&
  {Tammann}}]{BST1985}
{Binggeli} B., {Sandage} A., {Tammann} G.~A., 1985, \aj, 90, 1681

\bibitem[{{Binney} \& {Merrifield}(1998)}]{BinneyMerrifield1998}
{Binney} J., {Merrifield} M., 1998, {Galactic astronomy}, {Binney, J.~\&
  Merrifield, M.}, ed.

\bibitem[{{Blanton} {et~al.}(2003){Blanton}, {Hogg}, {Bahcall}, {Brinkmann},
  {Britton}, {Connolly}, {Csabai}, {Fukugita}, {Loveday}, {Meiksin}, {Munn},
  {Nichol}, {Okamura}, {Quinn}, {Schneider}, {Shimasaku}, {Strauss}, {Tegmark},
  {Vogeley}, \& {Weinberg}}]{Blanton+2003}
{Blanton} M.~R., {Hogg} D.~W., {Bahcall} N.~A., {Brinkmann} J., {Britton} M.,
  {Connolly} A.~J., {Csabai} I., {Fukugita} M., {Loveday} J., {Meiksin} A.,
  {Munn} J.~A., {Nichol} R.~C., {Okamura} S., {Quinn} T., {Schneider} D.~P.,
  {Shimasaku} K., {Strauss} M.~A., {Tegmark} M., {Vogeley} M.~S., {Weinberg}
  D.~H., 2003, \apj, 592, 819

\bibitem[{{Buzzoni}(2005)}]{Buzzoni2005}
{Buzzoni} A., 2005, \mnras, 361, 725

\bibitem[{{Cole} {et~al.}(2001){Cole}, {Norberg}, {Baugh}, {Frenk},
  {Bland-Hawthorn}, {Bridges}, {Cannon}, {Colless}, {Collins}, {Couch},
  {Cross}, {Dalton}, {De Propris}, {Driver}, {Efstathiou}, {Ellis},
  {Glazebrook}, {Jackson}, {Lahav}, {Lewis}, {Lumsden}, {Maddox}, {Madgwick},
  {Peacock}, {Peterson}, {Sutherland}, \& {Taylor}}]{Cole+2001}
{Cole} S., {Norberg} P., {Baugh} C.~M., {Frenk} C.~S., {Bland-Hawthorn} J.,
  {Bridges} T., {Cannon} R., {Colless} M., {Collins} C., {Couch} W., {Cross}
  N., {Dalton} G., {De Propris} R., {Driver} S.~P., {Efstathiou} G., {Ellis}
  R.~S., {Glazebrook} K., {Jackson} C., {Lahav} O., {Lewis} I., {Lumsden} S.,
  {Maddox} S., {Madgwick} D., {Peacock} J.~A., {Peterson} B.~A., {Sutherland}
  W., {Taylor} K., 2001, \mnras, 326, 255

\bibitem[{{Colless} {et~al.}(2001){Colless}, {Dalton}, {Maddox}, {Sutherland},
  {Norberg}, {Cole}, {Bland-Hawthorn}, {Bridges}, {Cannon}, {Collins}, {Couch},
  {Cross}, {Deeley}, {De Propris}, {Driver}, {Efstathiou}, {Ellis}, {Frenk},
  {Glazebrook}, {Jackson}, {Lahav}, {Lewis}, {Lumsden}, {Madgwick}, {Peacock},
  {Peterson}, {Price}, {Seaborne}, \& {Taylor}}]{2dF}
{Colless} M., {Dalton} G., {Maddox} S., {Sutherland} W., {Norberg} P., {Cole}
  S., {Bland-Hawthorn} J., {Bridges} T., {Cannon} R., {Collins} C., {Couch} W.,
  {Cross} N., {Deeley} K., {De Propris} R., {Driver} S.~P., {Efstathiou} G.,
  {Ellis} R.~S., {Frenk} C.~S., {Glazebrook} K., {Jackson} C., {Lahav} O.,
  {Lewis} I., {Lumsden} S., {Madgwick} D., {Peacock} J.~A., {Peterson} B.~A.,
  {Price} I., {Seaborne} M., {Taylor} K., 2001, \mnras, 328, 1039

\bibitem[{{Crook} {et~al.}(2007){Crook}, {Huchra}, {Martimbeau}, {Masters},
  {Jarrett}, \& {Macri}}]{Crook+2007}
{Crook} A.~C., {Huchra} J.~P., {Martimbeau} N., {Masters} K.~L., {Jarrett} T.,
  {Macri} L.~M., 2007, \apj, 655, 790

\bibitem[{{de Vaucouleurs} {et~al.}(1976){de Vaucouleurs}, {de Vaucouleurs}, \&
  {Corwin}}]{RC2}
{de Vaucouleurs} G., {de Vaucouleurs} A., {Corwin} J.~R., 1976, in Second
  reference catalogue of bright galaxies, 1976, Austin: University of Texas
  Press., pp. 0--+

\bibitem[{{Desai} {et~al.}(2004){Desai}, {Dalcanton}, {Mayer}, {Reed}, {Quinn},
  \& {Governato}}]{Desai+2004}
{Desai} V., {Dalcanton} J.~J., {Mayer} L., {Reed} D., {Quinn} T., {Governato}
  F., 2004, \mnras, 351, 265

\bibitem[{{D{\'{\i}}az-Gim{\'e}nez} {et~al.}(2008){D{\'{\i}}az-Gim{\'e}nez},
  {Muriel}, \& {Mendes de Oliveira}}]{DiazGimenez+2008}
{D{\'{\i}}az-Gim{\'e}nez} E., {Muriel} H., {Mendes de Oliveira} C., 2008, \aap,
  490, 965

\bibitem[{{Djorgovski} {et~al.}(1995){Djorgovski}, {Soifer}, {Pahre}, {Larkin},
  {Smith}, {Neugebauer}, {Smail}, {Matthews}, {Hogg}, {Blandford}, {Cohen},
  {Harrison}, \& {Nelson}}]{Djorgovski+1995}
{Djorgovski} S., {Soifer} B.~T., {Pahre} M.~A., {Larkin} J.~E., {Smith} J.~D.,
  {Neugebauer} G., {Smail} I., {Matthews} K., {Hogg} D.~W., {Blandford} R.~D.,
  {Cohen} J., {Harrison} W., {Nelson} J., 1995, \apjl, 438, L13

\bibitem[{{Eigenthaler} \& {Zeilinger}(2010)}]{EigenthalerZeilinger2010}
{Eigenthaler} P., {Zeilinger} W.~W., 2010, \aap, 511, A12+

\bibitem[{{Ferguson}(1989)}]{Ferguson1989}
{Ferguson} H.~C., 1989, \aj, 98, 367

\bibitem[{{Ferguson} \& {Sandage}(1990)}]{FergusonSandage1989}
{Ferguson} H.~C., {Sandage} A., 1990, \aj, 100, 1

\bibitem[{{Fukugita} \& {Peebles}(2004)}]{FukugitaPeebles2004}
{Fukugita} M., {Peebles} P.~J.~E., 2004, \apj, 616, 643

\bibitem[{{Fukugita} {et~al.}(1995){Fukugita}, {Shimasaku}, \&
  {Ichikawa}}]{Fukugita+1995}
{Fukugita} M., {Shimasaku} K., {Ichikawa} T., 1995, \pasp, 107, 945

\bibitem[{{Giovanelli} {et~al.}(2005){Giovanelli}, {Haynes}, {Kent},
  {Perillat}, {Saintonge}, {Brosch}, {Catinella}, {Hoffman}, {Stierwalt},
  {Spekkens}, {Lerner}, {Masters}, {Momjian}, {Rosenberg}, {Springob},
  {Boselli}, {Charmandaris}, {Darling}, {Davies}, {Garcia Lambas}, {Gavazzi},
  {Giovanardi}, {Hardy}, {Hunt}, {Iovino}, {Karachentsev}, {Karachentseva},
  {Koopmann}, {Marinoni}, {Minchin}, {Muller}, {Putman}, {Pantoja}, {Salzer},
  {Scodeggio}, {Skillman}, {Solanes}, {Valotto}, {van Driel}, \& {van
  Zee}}]{ALFALFA}
{Giovanelli} R., {Haynes} M.~P., {Kent} B.~R., {Perillat} P., {Saintonge} A.,
  {Brosch} N., {Catinella} B., {Hoffman} G.~L., {Stierwalt} S., {Spekkens} K.,
  {Lerner} M.~S., {Masters} K.~L., {Momjian} E., {Rosenberg} J.~L., {Springob}
  C.~M., {Boselli} A., {Charmandaris} V., {Darling} J.~K., {Davies} J., {Garcia
  Lambas} D., {Gavazzi} G., {Giovanardi} C., {Hardy} E., {Hunt} L.~K., {Iovino}
  A., {Karachentsev} I.~D., {Karachentseva} V.~E., {Koopmann} R.~A., {Marinoni}
  C., {Minchin} R., {Muller} E., {Putman} M., {Pantoja} C., {Salzer} J.~J.,
  {Scodeggio} M., {Skillman} E., {Solanes} J.~M., {Valotto} C., {van Driel} W.,
  {van Zee} L., 2005, \aj, 130, 2598

\bibitem[{{Gott} {et~al.}(2005){Gott}, {Juri{\'c}}, {Schlegel}, {Hoyle},
  {Vogeley}, {Tegmark}, {Bahcall}, \& {Brinkmann}}]{Gott+2005}
{Gott} III J.~R., {Juri{\'c}} M., {Schlegel} D., {Hoyle} F., {Vogeley} M.,
  {Tegmark} M., {Bahcall} N., {Brinkmann} J., 2005, \apj, 624, 463

\bibitem[{{Heisler} {et~al.}(1985){Heisler}, {Tremaine}, \&
  {Bahcall}}]{Heisler+1985}
{Heisler} J., {Tremaine} S., {Bahcall} J.~N., 1985, \apj, 298, 8

\bibitem[{{Hoffman} {et~al.}(1980){Hoffman}, {Olson}, \&
  {Salpeter}}]{Hoffman+1980}
{Hoffman} G.~L., {Olson} D.~W., {Salpeter} E.~E., 1980, \apj, 242, 861

\bibitem[{{Huang} {et~al.}(2001){Huang}, {Thompson}, {K{\"u}mmel},
  {Meisenheimer}, {Wolf}, {Beckwith}, {Fockenbrock}, {Fried}, {Hippelein}, {von
  Kuhlmann}, {Phleps}, {R{\"o}ser}, \& {Thommes}}]{Huang+2001}
{Huang} J., {Thompson} D., {K{\"u}mmel} M.~W., {Meisenheimer} K., {Wolf} C.,
  {Beckwith} S.~V.~W., {Fockenbrock} R., {Fried} J.~W., {Hippelein} H., {von
  Kuhlmann} B., {Phleps} S., {R{\"o}ser} H., {Thommes} E., 2001, \aap, 368, 787

\bibitem[{{Huchra} \& {Geller}(1982)}]{HuchraGeller1982}
{Huchra} J.~P., {Geller} M.~J., 1982, \apj, 257, 423

\bibitem[{{Jarrett} {et~al.}(2000){Jarrett}, {Chester}, {Cutri}, {Schneider},
  {Skrutskie}, \& {Huchra}}]{2MASSX}
{Jarrett} T.~H., {Chester} T., {Cutri} R., {Schneider} S., {Skrutskie} M.,
  {Huchra} J.~P., 2000, \aj, 119, 2498

\bibitem[{{Jarrett} {et~al.}(2003){Jarrett}, {Chester}, {Cutri}, {Schneider},
  \& {Huchra}}]{2MASSAtlas}
{Jarrett} T.~H., {Chester} T., {Cutri} R., {Schneider} S.~E., {Huchra} J.~P.,
  2003, \aj, 125, 525

\bibitem[{{Jee} {et~al.}(2005){Jee}, {White}, {Ben{\'{\i}}tez}, {Ford},
  {Blakeslee}, {Rosati}, {Demarco}, \& {Illingworth}}]{Jee+2005}
{Jee} M.~J., {White} R.~L., {Ben{\'{\i}}tez} N., {Ford} H.~C., {Blakeslee}
  J.~P., {Rosati} P., {Demarco} R., {Illingworth} G.~D., 2005, \apj, 618, 46

\bibitem[{{Jones} {et~al.}(2006){Jones}, {Peterson}, {Colless}, \&
  {Saunders}}]{Jones+2006}
{Jones} D.~H., {Peterson} B.~A., {Colless} M., {Saunders} W., 2006, \mnras,
  369, 25

\bibitem[{{Jones} {et~al.}(2004){Jones}, {Saunders}, {Colless}, {Read},
  {Parker}, {Watson}, {Campbell}, {Burkey}, {Mauch}, {Moore}, {Hartley},
  {Cass}, {James}, {Russell}, {Fiegert}, {Dawe}, {Huchra}, {Jarrett}, {Lahav},
  {Lucey}, {Mamon}, {Proust}, {Sadler}, \& {Wakamatsu}}]{6dF}
{Jones} D.~H., {Saunders} W., {Colless} M., {Read} M.~A., {Parker} Q.~A.,
  {Watson} F.~G., {Campbell} L.~A., {Burkey} D., {Mauch} T., {Moore} L.,
  {Hartley} M., {Cass} P., {James} D., {Russell} K., {Fiegert} K., {Dawe} J.,
  {Huchra} J., {Jarrett} T., {Lahav} O., {Lucey} J., {Mamon} G.~A., {Proust}
  D., {Sadler} E.~M., {Wakamatsu} K., 2004, \mnras, 355, 747

\bibitem[{{Jones} {et~al.}(2003){Jones}, {Ponman}, {Horton}, {Babul},
  {Ebeling}, \& {Burke}}]{Jones+2003}
{Jones} L.~R., {Ponman} T.~J., {Horton} A., {Babul} A., {Ebeling} H., {Burke}
  D.~J., 2003, \mnras, 343, 627

\bibitem[{{Karachentsev}(1994)}]{K1994}
{Karachentsev} I., 1994, Astronomical and Astrophysical Transactions, 6, 1

\bibitem[{{Karachentsev}(2005)}]{K2005}
{Karachentsev} I.~D., 2005, \aj, 129, 178

\bibitem[{{Karachentsev} \& {Karachentseva}(2004)}]{KK2004}
{Karachentsev} I.~D., {Karachentseva} V.~E., 2004, Astronomy Reports, 48, 267

\bibitem[{{Karachentsev} {et~al.}(2006){Karachentsev}, {Karachentseva}, \&
  {Huchtmeier}}]{KKH2006}
{Karachentsev} I.~D., {Karachentseva} V.~E., {Huchtmeier} W.~K., 2006, \aap,
  451, 817

\bibitem[{{Karachentsev} {et~al.}(2007){Karachentsev}, {Karachentseva}, \&
  {Huchtmeier}}]{KKH2007}
---, 2007, Astronomy Letters, 33, 512

\bibitem[{{Karachentsev} {et~al.}(2004){Karachentsev}, {Karachentseva},
  {Huchtmeier}, \& {Makarov}}]{CNG}
{Karachentsev} I.~D., {Karachentseva} V.~E., {Huchtmeier} W.~K., {Makarov}
  D.~I., 2004, \aj, 127, 2031

\bibitem[{{Karachentsev} \& {Kutkin}(2005)}]{KK2005}
{Karachentsev} I.~D., {Kutkin} A.~M., 2005, Astronomy Letters, 31, 299

\bibitem[{{Karachentsev} \& {Makarov}(2008)}]{Pairs}
{Karachentsev} I.~D., {Makarov} D.~I., 2008, Astrophysical Bulletin, 63, 299

\bibitem[{{Karachentsev} {et~al.}(2008){Karachentsev}, {Makarov},
  {Karachentseva}, \& {Melnik}}]{KMKM2008}
{Karachentsev} I.~D., {Makarov} D.~I., {Karachentseva} V.~E., {Melnik} O.~V.,
  2008, Astronomy Letters, 34, 832

\bibitem[{{Karachentsev} {et~al.}(2009){Karachentsev}, {Makarov},
  {Karachentseva}, \& {Melnyk}}]{KMKM2009}
{Karachentsev} I.~D., {Makarov} D.~I., {Karachentseva} V.~E., {Melnyk} O.~V.,
  2009, ArXiv e-prints

\bibitem[{{Karachentsev} \& {Nasonova}(2010)}]{KN2010}
{Karachentsev} I.~D., {Nasonova} O.~G., 2010, \mnras, 405, 1075

\bibitem[{{Karachentseva} \& {Karachentsev}(1998)}]{KK1998}
{Karachentseva} V.~E., {Karachentsev} I.~D., 1998, \aaps, 127, 409

\bibitem[{{Karachentseva} \& {Karachentsev}(2000)}]{KK2000}
---, 2000, \aaps, 146, 359

\bibitem[{{Karachentseva} {et~al.}(2011){Karachentseva}, {Melnyk},
  {Karachentsev}, \& {Makarov}}]{Karachentseva+2011}
{Karachentseva} V.~E., {Melnyk} O.~V., {Karachentsev} I.~D., {Makarov} D.~I.,
  2011, in preparation

\bibitem[{{Kochanek} {et~al.}(2001){Kochanek}, {Pahre}, {Falco}, {Huchra},
  {Mader}, {Jarrett}, {Chester}, {Cutri}, \& {Schneider}}]{Kochanek+2001}
{Kochanek} C.~S., {Pahre} M.~A., {Falco} E.~E., {Huchra} J.~P., {Mader} J.,
  {Jarrett} T.~H., {Chester} T., {Cutri} R., {Schneider} S.~E., 2001, \apj,
  560, 566

\bibitem[{{Magtesyan}(1988)}]{Magtesyan1988}
{Magtesyan} A.~P., 1988, Astrophysics, 28, 150

\bibitem[{{Mahdavi} {et~al.}(2005){Mahdavi}, {Trentham}, \&
  {Tully}}]{Mahdavi+2005}
{Mahdavi} A., {Trentham} N., {Tully} R.~B., 2005, \aj, 130, 1502

\bibitem[{{Maia} {et~al.}(1989){Maia}, {da Costa}, \& {Latham}}]{MdCL1989}
{Maia} M.~A.~G., {da Costa} L.~N., {Latham} D.~W., 1989, \apjs, 69, 809

\bibitem[{{Makarov} \& {Karachentsev}(2000)}]{MK2000}
{Makarov} D.~I., {Karachentsev} I.~D., 2000, in Astronomical Society of the
  Pacific Conference Series, Vol. 209, IAU Colloq. 174: Small Galaxy Groups,
  {M.~J.~Valtonen \& C.~Flynn}, ed., pp. 40--+

\bibitem[{{Makarov} \& {Karachentsev}(2009)}]{Triplets}
---, 2009, Astrophysical Bulletin, 64, 24

\bibitem[{{Mandelbaum} {et~al.}(2006){Mandelbaum}, {Seljak}, {Kauffmann},
  {Hirata}, \& {Brinkmann}}]{Mandelbaum+2006}
{Mandelbaum} R., {Seljak} U., {Kauffmann} G., {Hirata} C.~M., {Brinkmann} J.,
  2006, \mnras, 368, 715

\bibitem[{{Materne}(1978)}]{Materne1978}
{Materne} J., 1978, \aap, 63, 401

\bibitem[{{Materne}(1979)}]{Materne1979}
---, 1979, \aap, 74, 235

\bibitem[{{McLaughlin}(1999)}]{McLaughlin1999}
{McLaughlin} D.~E., 1999, \apjl, 512, L9

\bibitem[{{Mieske} {et~al.}(2007){Mieske}, {Hilker}, {Infante}, \& {Mendes de
  Oliveira}}]{Mieske+2007}
{Mieske} S., {Hilker} M., {Infante} L., {Mendes de Oliveira} C., 2007, \aap,
  463, 503

\bibitem[{{Nasonova} {et~al.}(2011){Nasonova}, {Peirani}, \& {de Freitas
  Pacheco}}]{Nasonova+2011}
{Nasonova} O.~G., {Peirani} S., {de Freitas Pacheco} J.~A., 2011, in
  preparation

\bibitem[{{Natarajan} \& {Springel}(2004)}]{NS2004}
{Natarajan} P., {Springel} V., 2004, \apjl, 617, L13

\bibitem[{{P{\'a}pai} \& {Szapudi}(2010)}]{PS2010}
{P{\'a}pai} P., {Szapudi} I., 2010, ArXiv e-prints

\bibitem[{{Paturel} {et~al.}(2003){Paturel}, {Petit}, {Prugniel}, {Theureau},
  {Rousseau}, {Brouty}, {Dubois}, \& {Cambr{\'e}sy}}]{HyperLEDA}
{Paturel} G., {Petit} C., {Prugniel} P., {Theureau} G., {Rousseau} J., {Brouty}
  M., {Dubois} P., {Cambr{\'e}sy} L., 2003, \aap, 412, 45

\bibitem[{{Sandage}(1986)}]{Sandage1986}
{Sandage} A., 1986, \apj, 307, 1

\bibitem[{{Schechter}(1976)}]{Schechter1976}
{Schechter} P., 1976, \apj, 203, 297

\bibitem[{{Schlegel} {et~al.}(1998){Schlegel}, {Finkbeiner}, \&
  {Davis}}]{Dustmap}
{Schlegel} D.~J., {Finkbeiner} D.~P., {Davis} M., 1998, \apj, 500, 525

\bibitem[{{Spergel} {et~al.}(2007){Spergel}, {Bean}, {Dor{\'e}}, {Nolta},
  {Bennett}, {Dunkley}, {Hinshaw}, {Jarosik}, {Komatsu}, {Page}, {Peiris},
  {Verde}, {Halpern}, {Hill}, {Kogut}, {Limon}, {Meyer}, {Odegard}, {Tucker},
  {Weiland}, {Wollack}, \& {Wright}}]{Spergel+2007}
{Spergel} D.~N., {Bean} R., {Dor{\'e}} O., {Nolta} M.~R., {Bennett} C.~L.,
  {Dunkley} J., {Hinshaw} G., {Jarosik} N., {Komatsu} E., {Page} L., {Peiris}
  H.~V., {Verde} L., {Halpern} M., {Hill} R.~S., {Kogut} A., {Limon} M.,
  {Meyer} S.~S., {Odegard} N., {Tucker} G.~S., {Weiland} J.~L., {Wollack} E.,
  {Wright} E.~L., 2007, \apjs, 170, 377

\bibitem[{{Springob} {et~al.}(2007){Springob}, {Masters}, {Haynes},
  {Giovanelli}, \& {Marinoni}}]{Springob+2007}
{Springob} C.~M., {Masters} K.~L., {Haynes} M.~P., {Giovanelli} R., {Marinoni}
  C., 2007, \apjs, 172, 599

\bibitem[{{Tonry} {et~al.}(2000){Tonry}, {Blakeslee}, {Ajhar}, \&
  {Dressler}}]{Tonry+2000}
{Tonry} J.~L., {Blakeslee} J.~P., {Ajhar} E.~A., {Dressler} A., 2000, \apj,
  530, 625

\bibitem[{{Tonry} {et~al.}(2001){Tonry}, {Dressler}, {Blakeslee}, {Ajhar},
  {Fletcher}, {Luppino}, {Metzger}, \& {Moore}}]{Tonry+2001}
{Tonry} J.~L., {Dressler} A., {Blakeslee} J.~P., {Ajhar} E.~A., {Fletcher}
  A.~B., {Luppino} G.~A., {Metzger} M.~R., {Moore} C.~B., 2001, \apj, 546, 681

\bibitem[{{Totani} {et~al.}(2001){Totani}, {Yoshii}, {Maihara}, {Iwamuro}, \&
  {Motohara}}]{Totani+2001}
{Totani} T., {Yoshii} Y., {Maihara} T., {Iwamuro} F., {Motohara} K., 2001,
  \apj, 559, 592

\bibitem[{{Trentham} \& {Tully}(2009)}]{TT2009}
{Trentham} N., {Tully} R.~B., 2009, \mnras, 398, 722

\bibitem[{{Tully}(1987)}]{Tully1987}
{Tully} R.~B., 1987, \apj, 321, 280

\bibitem[{{Tully}(1988)}]{NGC}
---, 1988, {Nearby galaxies catalog}, {Tully, R.~B.}, ed.

\bibitem[{{Tully} \& {Fisher}(1977)}]{TullyFisher}
{Tully} R.~B., {Fisher} J.~R., 1977, \aap, 54, 661

\bibitem[{{Tully} \& {Shaya}(1984)}]{TS1984}
{Tully} R.~B., {Shaya} E.~J., 1984, \apj, 281, 31

\bibitem[{{Tully} \& {Trentham}(2008)}]{TT2008}
{Tully} R.~B., {Trentham} N., 2008, \aj, 135, 1488

\bibitem[{{Vennik}(1984)}]{Vennik1984}
{Vennik} J., 1984, Tartu Astrofuusika Observatoorium Teated, 73, 1

\bibitem[{{Vennik}(1987)}]{Vennik1987}
---, 1987, Dissertation. Tartu

\bibitem[{{von Benda-Beckmann} {et~al.}(2008){von Benda-Beckmann}, {D'Onghia},
  {Gottl{\"o}ber}, {Hoeft}, {Khalatyan}, {Klypin}, \&
  {M{\"u}ller}}]{vonBendaBeckmann+2008}
{von Benda-Beckmann} A.~M., {D'Onghia} E., {Gottl{\"o}ber} S., {Hoeft} M.,
  {Khalatyan} A., {Klypin} A., {M{\"u}ller} V., 2008, \mnras, 386, 2345

\bibitem[{{Zandivarez} {et~al.}(2006){Zandivarez}, {Mart{\'{\i}}nez}, \&
  {Merch{\'a}n}}]{ZMM2006}
{Zandivarez} A., {Mart{\'{\i}}nez} H.~J., {Merch{\'a}n} M.~E., 2006, \apj, 650,
  137

\end{thebibliography}

\label{lastpage}

\end{document}